\documentclass{elsart}

\usepackage{graphicx}

\usepackage{amssymb}
\usepackage{psfrag}

\newcommand\be{\begin{equation}}
\newcommand\ee{\end{equation}}
\newcommand{\bea}{\begin{eqnarray}}
\newcommand{\eea}{\end{eqnarray}}

\newcommand\pdv[2]{\frac{\partial#1}{\partial#2}}
\newcommand{\Dv}[2]{\frac{D#1}{D#2}}
\newcommand\bld[1]{\mbox{\boldmath $#1$}}
\newcommand{\bnabla}{\bld{\nabla}}
\newcommand{\bv}{\bld{v}}
\newcommand{\bk}{\bld{k}}
\newcommand{\bx}{\bld{x}}

\newcommand\cm{{\rm\,cm}}

\newcommand\gm{{\rm\,g}}

\newcommand\K{{\rm\,K}}
\newcommand\nbu{N_{Bu}}
\newcommand\nbo{N_{Bo}}
\newcommand\txi{\tilde{\xi}}
\newcommand\sj{\Sigma_j}

\begin{document}

\begin{frontmatter}



\title{Numerical Tests and Properties of Waves in Radiating Fluids}


\author[a1,a2]{Bryan M. Johnson}
\author[a1,a3]{\& Richard I. Klein}

\address[a1]{Lawrence Livermore National Laboratory, AX Division, 7000 East Avenue, Livermore, CA 94550}
\address[a2]{johnson359@llnl.gov}
\address[a3]{Department of Astronomy, University of California, Berkeley 601 Campbell Hall, Berkeley, CA 94720}

\begin{abstract}
We discuss the properties of an analytical solution for waves in radiating fluids, with a view towards its implementation as a quantitative test of radiation hydrodynamics codes. A homogeneous radiating fluid in local thermodynamic equilibrium is periodically driven at the boundary of a one-dimensional domain, and the solution describes the propagation of the waves thus excited. Two modes are excited for a given driving frequency, generally referred to as a radiative acoustic wave and a radiative diffusion wave. While the analytical solution is well known, several features are highlighted here that require care during its numerical implementation. We compare the solution in a wide range of parameter space to a numerical integration with a Lagrangian radiation hydrodynamics code. Our most significant observation is that flux-limited diffusion does not preserve causality for waves on a homogeneous background.
\end{abstract}

\end{frontmatter}

\section{Introduction}
\label{}

Analytical solutions for radiation hydrodynamics are difficult to obtain due to the complexity of the equations but are a powerful tool for testing complex multi-physics codes. One of the most useful simplifying assumptions for any set of nonlinear differential equations is to set up an equilibrium state and analyze small departures from that equilibrium. Since this approach retains most of the terms in the equations, it not only provides valuable physical insight but also serves as a comprehensive and sensitive test of numerical algorithms. Many perturbation studies of radiation hydrodynamics have been performed; we follow closely Mihalas \& Mihalas \cite{mm83, mm84} and Bogdan et al. \cite{bog96}, and refer the reader there for additional references. Despite the straightforward application of perturbation theory to the equations of radiation hydrodynamics, however, there appear to be few numerical tests of this type of solution in the literature (reference \cite{ts01} is one example). Our goal here is to conduct a systematic comparison of such a solution with a numerical algorithm in a wide range of parameter space. The code that we use for comparison is the Lagrangian radiation hydrodynamics code Kull \cite{kull00}. We begin in \S\ref{AE} with an overview of our assumptions and the form the equations of radiation hydrodynamics take under these assumptions. The failure of flux-limited diffusion to capture free-streaming radiation waves is highlighted in \S\ref{FLD}. We discuss the analytical solution in \S\ref{AS} and compare our results with previous work in \S\ref{PW}. Numerical results are given in \S\ref{NR} and we summarize in \S\ref{SD}. 

\section{Assumptions and Equations}
\label{AE}

We investigate perturbations from an equilibrium state of constant density and temperature with zero velocity and zero radiation flux. In addition to dropping terms that are higher than linear order in the perturbation amplitude, we further simplify the equations by making the following standard assumptions: 1) the material fluid is an ideal gas, 2) the material and radiation are in local thermodynamic equilibrium (LTE)\footnote{Note that a common temperature for the material and radiation only applies to the equilibrium state; the material and radiation temperature perturbations are allowed to differ.}, 3) the opacity is independent of frequency, and 4) scattering is negligible. The common assumption of an opacity that is also independent of temperature and density is not strictly necessary for a perturbation analysis; one can easily show that variations in the opacity due to density and temperature perturbations give rise to terms that are higher than linear order in the analysis.

The above assumptions must be supplemented with a prescription for the configuration of the radiation field. One approach is to solve the radiation transport equation directly, making some assumption for the angular distribution of the radiation \cite{bog96}. Alternatively, one can calculate angular moments of the transport equation and invoke a prescription for closing the moment equations. A commonly employed closure scheme is the Eddington approximation, which assumes that the radiation stress is isotropic and given by ${\bf P} = (E/3){\bf I}$, where $E$ is the radiation energy density and ${\bf I}$ is the unit tensor. This is the approach taken by, for example, Mihalas \& Mihalas \cite{mm83, mm84}.

Both of these approaches are numerically expensive, however, due to the large dynamic range between the length and time scales of the material and radiation. The disparity in time scales can be alleviated somewhat by invoking the diffusion approximation, which assumes that the time dependence of the radiation flux is negligible. Since this can result in a superluminal flux of radiation energy, numerical calculations typically employ some type of flux-limited diffusion, which gives one the computational advantages of the diffusion approximation while preventing the flux from becoming unphysical.

As we discuss in the following section, however, all flux limiters reduce to the diffusion limit for linear perturbations. As a result, our numerical calculations are in the diffusion limit, although we discuss the analytical solution under the Eddington approximation for comparison with previous work. The equations of radiation hydrodynamics under the Eddington approximation in a frame comoving with the fluid\footnote{Since our perturbation analysis implies a constant opacity, the results are independent of the choice of reference frame.} are
\be\label{EQ1}
\Dv{\rho}{t} = -\rho \bnabla \cdot \bv,
\ee
\be\label{EQ2}
\rho \Dv{\bv}{t} = -\bnabla p + \frac{\chi}{c} \bld{F},
\ee
\be\label{EQ3}
\Dv{p}{t} = -\gamma p \bnabla \cdot \bv  + c \chi (\gamma - 1)\left(E - a_BT^4\right),
\ee
\be\label{EQ4}
\Dv{E}{t} = -\bnabla \cdot \bld{F} - \frac{4}{3} E \bnabla \cdot \bv + c \chi \left(a_BT^4 - E\right),
\ee
\be\label{EQ5}
\frac{1}{c}\Dv{\bld{F}}{t} = -\frac{c}{3} \bnabla E -\frac{1}{c} \bld{F} \bnabla \cdot \bv - \chi \bld{F},
\ee
where $\rho$, $p$ and $T$ are the material density, pressure and temperature, respectively, $\bv$ is the fluid velocity, $\bld{F}$ is the radiation flux, $a_B$ is the radiation constant, $c$ is the speed of light, and $\chi$ is the absorption opacity in units of inverse length.

The perturbed form of the above equations is
\be\label{CONT}
\pdv{}{t} \left(\frac{\delta \rho}{\rho_0}\right) + \bnabla \cdot \delta \bv  = 0,
\ee
\be\label{MM}
\frac{\gamma}{a}\pdv{}{t} \left(\frac{\delta \bv}{a}\right) + \bnabla\left(\frac{\delta p}{p_0}\right) - \frac{16\gamma r \chi}{c(\gamma - 1)} \frac{\delta \bld{F}}{4E_0} = 0,
\ee
\be\label{ME}
\pdv{}{t} \left(\frac{\delta p}{p_0} - \gamma \frac{\delta \rho}{\rho_0}\right) - 16\gamma r \chi c \left(\frac{\delta T_r}{T_0} - \frac{\delta T}{T_0}\right) = 0,
\ee
\be\label{RE}
\pdv{}{t}\left(\frac{\delta T_r}{T_0}\right) + \bnabla \cdot \left(\frac{\delta \bld{F}}{4E_0} + \frac{1}{3} \delta \bv\right) - c \chi \left(\frac{\delta T}{T_0} - \frac{\delta T_r}{T_0}\right) = 0,
\ee
\be\label{RM}
\pdv{}{t}\left(\frac{\delta \bld{F}}{4E_0} + \frac{1}{3} \delta \bv\right) + \frac{1}{3}c^2\bnabla \left(\frac{\delta T_r}{T_0}\right) + c\chi \frac{\delta \bld{F}}{4E_0} = 0,
\ee
where $T_r$ is the radiation temperature defined via $E = a_BT_r^4$ ($E_0 = a_B T_0^4$), $a = (\gamma p_0/\rho_0)^{1/2}$ is the material sound speed, and the dimensionless ratio 
\be
r \equiv \frac{(\gamma - 1)a_BT_0^4}{4\gamma p_0}
\ee
governs the coupling between the radiation and the material; it is proportional to the ratio of their energy densities. The subscript zero denotes an equilibrium quantity. For an ideal gas, the material pressure perturbation is given by
\be\label{IG}
\frac{\delta p}{p_0} = \frac{\delta T}{T_0} + \frac{\delta \rho}{\rho_0}.
\ee
Under the diffusion approximation, the perturbed radiation energy and momentum equations (\ref{RE}) and (\ref{RM}) reduce to
\be\label{REDIFF}
\pdv{}{t}\left(\frac{\delta T_r}{T_0}\right) - \frac{c}{3\chi} \nabla^2 \left(\frac{\delta T_r}{T_0}\right) + \frac{1}{3} \bnabla \cdot \delta \bv - c \chi \left(\frac{\delta T}{T_0} - \frac{\delta T_r}{T_0}\right) = 0
\ee
and
\be
\frac{\delta \bld{F}}{4E_0} = -\frac{c}{3\chi}\bnabla \left(\frac{\delta T_r}{T_0}\right).
\ee

\section{Breakdown of Flux-Limited Diffusion}
\label{FLD}

The radiation flux under the flux-limited diffusion approximation has the form (in the comoving frame)
\be\label{FLUX}
\bld{F} = -\frac{c\lambda}{\chi}\bnabla E,
\ee
where $\lambda$ is the flux limiter, designed to reduce to $1/3$ in optically thick regions and $\chi E/|\bnabla E|$ in optically thin regions. Flux limiters can take various forms but are generally nonlinear functions of
\be\label{FL}
R \equiv \frac{|\bnabla E|}{\chi E}.
\ee
The diffusion limit corresponds to $R \ll 1$. For an inhomogeneous medium such as an atmosphere, $|\bnabla E| \sim E/L$, where $L$ is the characteristic length scale of the medium. The magnitude of $R$ then depends only upon the optical depth $\tau_L \equiv \chi L$. In the case of small perturbations on a homogeneous background, however,
\be
R \sim \frac{k}{\chi} \frac{\delta E}{E},
\ee
where $k$ is the wave number of the perturbation, so that $R$ depends upon both a perturbation optical depth $\tau_k \equiv \chi/k$ and a perturbation amplitude. A linear solution clearly requires $R \ll 1$, even when $\tau^{-1}_k \gg 1$. All flux limiters therefore operate in the diffusion limit for linear perturbations on a homogeneous background with zero mean flux. {\it In regions of the flow where the diffusion approximation breaks down, this results in superluminal propagation speeds.}

This behavior, while somewhat counter-intuitive, is to be expected since flux-limited diffusion is not designed to follow wave fronts. The flux for large $R$ (the free-streaming limit) is designed to reduce to $\bld{F} = cE\hat{\bld{n}}$, where $\hat{\bld{n}}$ is a unit vector in the direction of propagation; this constant flux can be viewed as a phase-averaged wave amplitude multiplied by a group velocity (i.e., a phase-averaged Poynting flux). Capturing the wave oscillations themselves clearly requires retaining the relevant time dependent and gradient terms in the equations; for free-streaming radiation, these are precisely the terms that are neglected in flux-limited diffusion (see, e.g., Levermore \& Pomraning \cite{lp81} equation [14]).

\section{Analytical Solution}
\label{AS}

Perturbations on a homogeneous background are naturally decomposed in terms of Fourier modes, whose space-time dependence is $\exp(i\omega t - i \bk \cdot \bx)$. As discussed in Bogdan et al. \cite{bog96}, these plane wave solutions to the perturbation equations can be decoupled into modes parallel and perpendicular to the velocity. The transverse modes are akin to viscous shear modes in hydrodynamics and we will not discuss them further here. For a system driven at a constant frequency $\omega$, the longitudinal modes give rise to a fourth-order dispersion relation for the wave number:
\be\label{DR}
c_4 \, \tau_k^{-4} + c_2 \, \tau_k^{-2} + c_0 = 0,
\ee
where the coefficients take different forms depending upon the approximation being used. For details on the derivation of the results in this section, see Appendix \ref{EE}. 

\subsection{Eddington Approximation}\label{EL}

The coefficients of the dispersion relation (\ref{DR}) under the Eddington approximation are
\be\label{C4}
c_4 = 1 - i16 r\tau_c,
\ee
\bea\label{C2}
c_2 = 3(1 + i\tau_c^{-1})^2 -\tau_a^{-2}(1 - i16\gamma r\tau_c) 
\nonumber \\
+ \; 16 r\left(5 + 3i\tau_c^{-1} + \frac{16\gamma r + i\tau_c^{-1}}{3[\gamma - 1]}\right),
\eea
\be\label{C0}
c_0 = -3\tau_a^{-2}\left(1 + i\tau_c^{-1} + 16\gamma r\right)\left(1 + i\tau_c^{-1} + \frac{16 r a^2}{3[\gamma - 1] c^2}\right),
\ee
where
\be
\tau_a \equiv \frac{a\chi}{\omega}
\ee
and
\be
\tau_c \equiv \frac{c\chi}{\omega}.
\ee

\begin{figure}
\psfrag{r}{$\beta^{-1}$}
\centering
\includegraphics[width=6.0in]{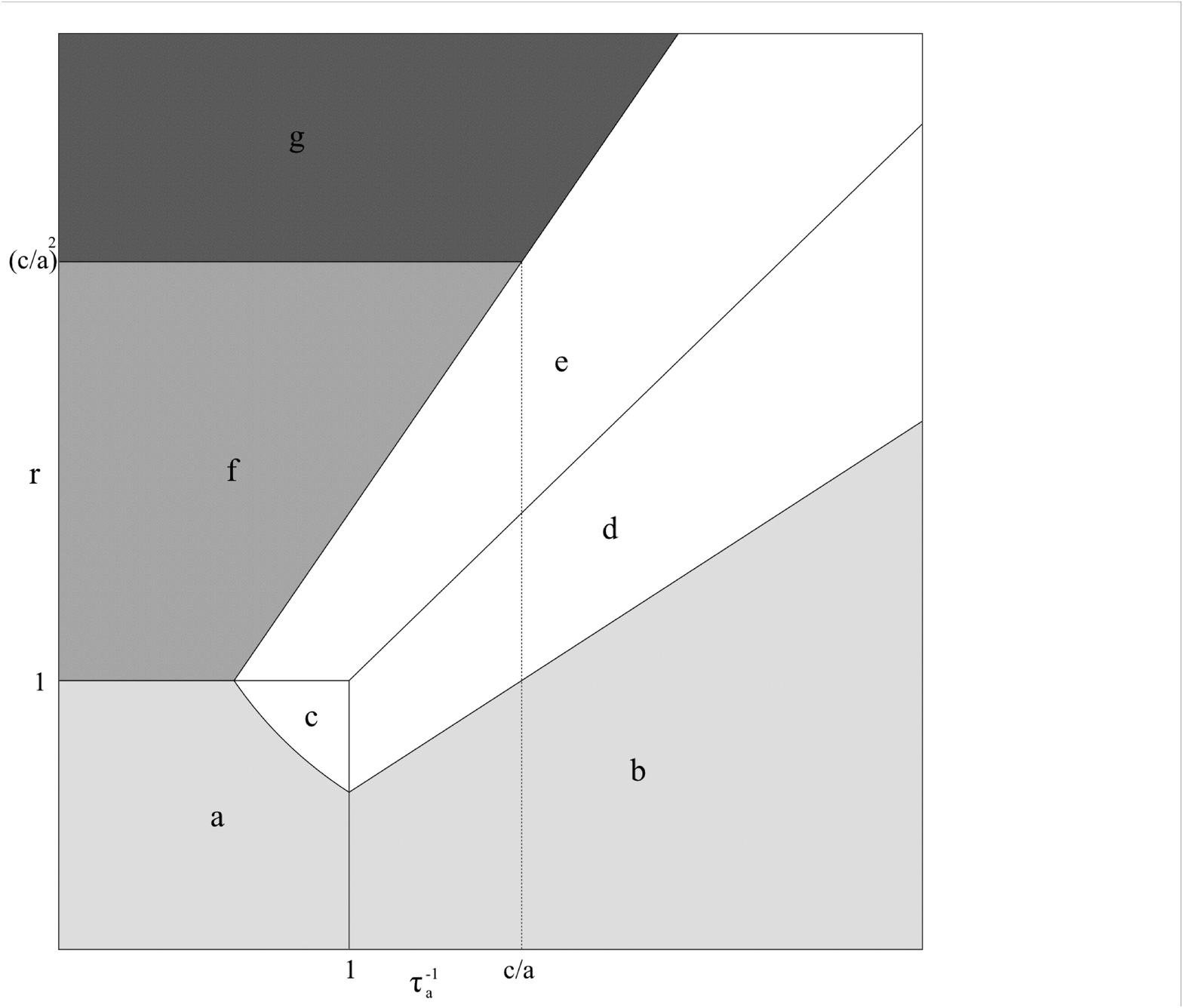}
\caption{Parameter space for the radiative acoustic mode under the Eddington approximation. The phase speeds are the material sound speed (regions a and b), the isothermal sound speed (regions c-e), the radiative sound speed (region f) and $c/\sqrt{3}$ (region g).}
\label{fra}
\end{figure}

\begin{figure}
\centering
\includegraphics[width=6.0in]{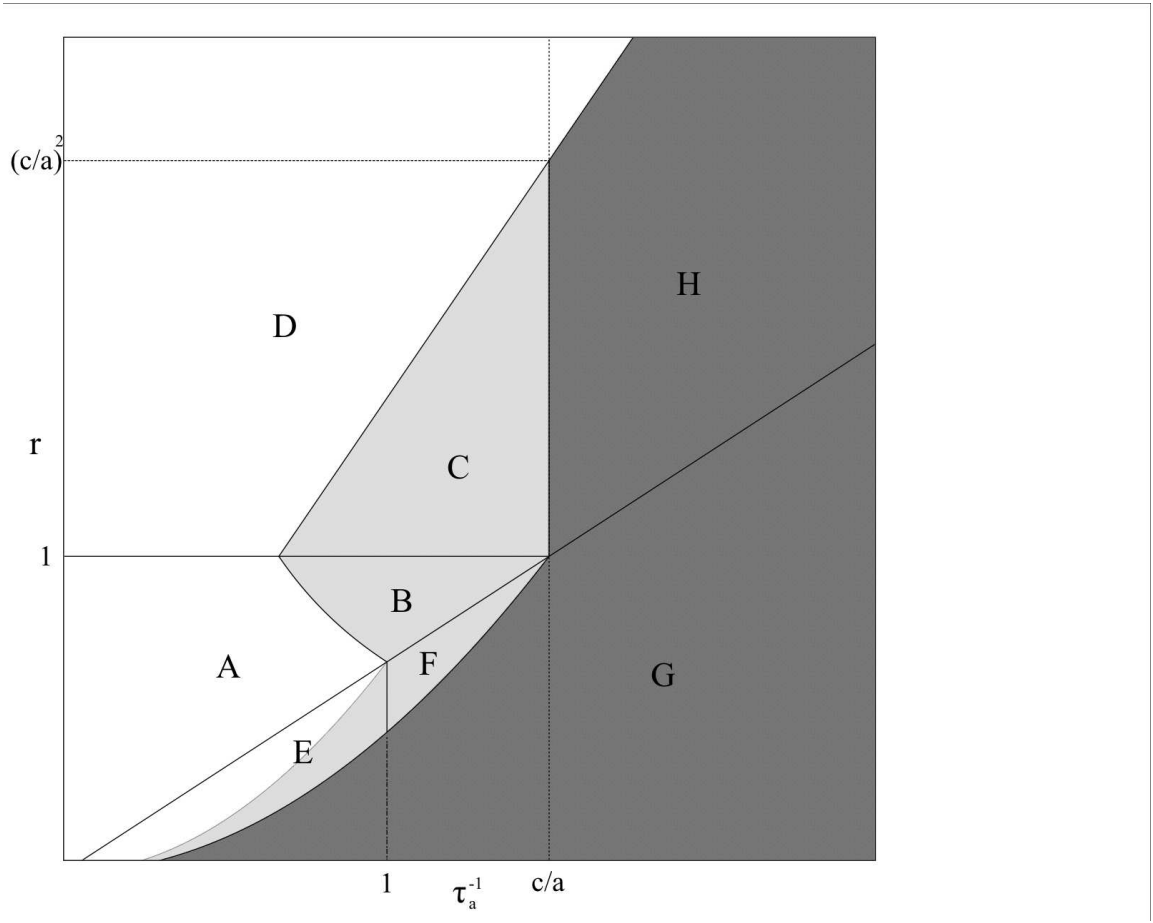}
\caption{Parameter space for the radiative diffusion mode under the Eddington approximation. The phase speeds are $v_p \ll a$ in the white region, $a < v_p < c$ in the light shaded region and $v_p = c\sqrt{3}$ in the dark shaded region.}
\label{frd}
\end{figure}

As discussed in Bogdan et al. \cite{bog96}, the solutions to the dispersion relation (\ref{DR}) have a simple form in most of parameter space. In Figures~\ref{fra} and \ref{frd} we reproduce Figure 4 of Bogdan et al. \cite{bog96} under the Eddington approximation;. The leading order solution in the regions defined by Figure~\ref{fra} for the radiative acoustic mode is
\be\label{KA}
k_a = \frac{\omega}{a}
\left\{
\begin{array}{cc}
1 - i8(\gamma-1)rc/(3 a\tau_a) & \;\;\; {\rm region \; a} \\ 
1 - i8(\gamma-1)r\tau_a c/a  & \;\;\; {\rm region \; b} \\
\sqrt{\gamma}\left(1 - i3 [\gamma-1] a\tau_a/[32\gamma^2 rc] \right)  & \;\;\; {\rm region \; c} \\
\sqrt{\gamma}\left(1 - ia[\gamma-1]/[32\gamma rc\tau_a] \right)  & \;\;\; {\rm region \; d} \\
\sqrt{\gamma}\left(1 - i8ra\tau_a/[3(\gamma-1)c] \right)  & \;\;\; {\rm region \; e} \\
\sqrt{9(\gamma-1)/(16r)}\left(1 - i3[\gamma-1] c/[32 r a \tau_a] \right)  & \;\;\; {\rm region \; f}  \\
\sqrt{3}(a/c)\left(1 - i9[\gamma-1]^2 c^3/[512r^2 a^3\tau_a]\right) & \;\;\; {\rm region \; g} \\
\end{array}
\right. 
\ee
These all have the form $\omega/k_a = v_p(1 + i\epsilon)$, where the phase velocity $v_p$ is $a$ in regions a and b, the isothermal sound speed $a/\sqrt{\gamma}$ in regions c-e, the radiative sound speed $\sqrt{4P/(3\rho)}$ in region f (where $P = E/3$ is the radiation pressure), and $c/\sqrt{3}$ in region g. Except near the region borders, the radiative acoustic wave is weakly damped (its damping length is much greater than its wavelength).

The leading order solution for the radiative diffusion mode (Figure~\ref{frd}) is
\be\label{KD}
\frac{k_d}{\chi} = 
\left\{
\begin{array}{cc}
\sqrt{3/(32 r\tau_c)} \; (1 - i) & \;\;\; {\rm region \; A} \\ 
\sqrt{3/(32\gamma r\tau_c)} \; (1 - i)  & \;\;\; {\rm region \; B} \\
\sqrt{3/(2\tau_c)} \; (1 - i) & \;\;\; {\rm region \; C} \\
\sqrt{8\gamma r/(3[\gamma-1]\tau_c)} \; (1 - i) & \;\;\; {\rm region \; D} \\
8\sqrt{3}r\tau_c\left(1 - i/[8 r \tau_c]\right) & \;\;\; {\rm region \; E} \\
8\sqrt{3}\gamma r\tau_c\left(1 - i /[8\gamma r \tau_c]\right) & \;\;\; {\rm region \; F} \\
(\sqrt{3}/\tau_c)\left(1 - i\tau_c\right) & \;\;\; {\rm region \; G} \\
(\sqrt{3}/\tau_c)\left(1 - i\tau_c/2\right) & \;\;\; {\rm region \; H} \\
\end{array}
\right. 
\ee
This wave is strongly damped everywhere except region H and the $\tau_c \ll 1$ portion of region G. In regions A-D the damping length is on the order of the perturbation wavelength, and in regions E and F and the $\tau_c \gg 1$ portion of region G it is much greater than a wavelength. The phase speeds are $v_p \ll a$ in regions A and D and a portion of region E, $a < v_p < c$ in regions B, C and F and a portion of region E and $v_p = c/\sqrt{3}$ in regions G and H.
 
\subsection{Diffusion Approximation}\label{DL}

Under the diffusion approximation, the solution for the radiative acoustic wave remains the same everywhere except region g, where it can be seen from expression (\ref{KA}) that its phase speed becomes superluminal.\footnote{This is also true for the dispersion relation of Mihalas \& Mihalas \cite{mm83, mm84}; see \S\ref{PW}.} The radiative sound speed $c_r \sim r^{1/2} a \rightarrow c$ as $r \rightarrow c^2/a^2$, which is the precisely the limit in which the final term in $c_0$ becomes important. As noted by Bogdan et al. \cite{bog96}, however, a consistent treatment of this region of parameter space would require relativistic physics, since the radiation energy density is greater than the rest mass energy density of the material.

The solution for the radiative diffusion wave remains the same everywhere except regions G and H. Region C extends into region H and the portion of region G for which $\tau_c \ll 1$, so that the mode remains diffusive rather than becoming free-streaming. Its phase speed is superluminal and increases without bound as the driving frequency is increased.\footnote{Since diffusive modes have $\omega \propto k^2$, their group speed is twice their phase speed.} Its damping length is on the order of a wave length. In the portion of region G for which $\tau_c \gg 1$, the solution is given by
\be
\frac{k_d}{\chi} = \frac{\sqrt{3}}{2\tau_c}\left(1 - i2\tau_c\right),
\ee
so that the mode is free-streaming with a phase speed (and group speed) $\sim 1.2c$ and a damping length half as long as under the Eddington approximation.

\section{Comparison with Previous Work}\label{PW}

\subsection{Mihalas \& Mihalas}

Our coefficients (\ref{C4})-(\ref{C0}) are somewhat different from those given by Mihalas \& Mihalas \cite{mm83, mm84} due to their neglect of the velocity dependent term in the radiation momentum equation (\ref{RM}).\footnote{This term arises due to the Doppler shift between the comoving and laboratory frames. Mihalas \& Mihalas \cite{mm84} refer to it as an acceleration term since in the comoving frame it appears as a time derivative of the velocity.} Comparison with expression (3.12) of Mihalas \& Mihalas \cite{mm83} reveals three differences: the term $3i\tau_c^{-1}$ in $c_2$ and the final term $\propto ra^2/c^2$ in $c_0$ are missing from the dispersion relation of Mihalas \& Mihalas \cite{mm83}, and they have an additional factor of $1 + i\tau_c^{-1}$ multiplying the final terms in parentheses in $c_2$ which cancels out in a self-consistent treatment. The only one of these differences that appears to be significant is the final term of $c_0$; as we discussed in \S\ref{DL}, this term is essential for limiting the phase speed to less than the speed of light at sufficiently large values of $r$.

\subsection{Bogdan, Knoelker, MacGregor \& Kim}

Bogdan et al. \cite{bog96} attempt to capture both the optically thick and optically thin regimes by explicitly calculating angular moments of the perturbed intensity, obtained directly from the perturbed transport equation. They obtain a transcendental equation for the wave number that reduces to a quadratic equation in both the optically thick and optically thin regimes. Remarkably, despite significant differences between the coefficients of their dispersion relation and ours,\footnote{This is due to the fact that the transcendental equation of Bogdan et al. \cite{bog96} must be expanded to $O(\tau_k^{-6})$ due to a cancellation at $O(\tau_k^0)$.} the solutions in all of the asymptotic regions (expressions [\ref{KA}] and [\ref{KD}]) are equivalent with the exception of region g and those regions for which the transcendental equation of Bogdan et al. \cite{bog96} does not reduce to a simple quadratic (our regions E-H). In order to perform angular integrals of the intensity, however, Bogdan et al. \cite{bog96} make the assumption that the velocity and material temperature perturbations are independent of angle. This does not appear to be a valid assumption in the optically thin limit, since coupling between the radiation and material should depend in that case upon the direction of the radiation.

\subsection{Lowrie, Morel \& Hittinger}

Lowrie et al. \cite{lmh99} analyze perturbations about a nonzero mean flow and focus on the initial value problem (solving for $\omega$ as a function of $k$). Under the Eddington approximation, the dispersion relation (\ref{DR}) is a fifth-order polynomial in $\omega$ that must be solved numerically. Both for simplicity and due to the fact that Kull does not currently currently support periodic boundary conditions, we have chosen to focus on the boundary value problem in this work. A periodic domain has its own advantages, however, and the initial value problem as a numerical test would not be plagued by the transients and reflections we have observed when conducting our numerical tests (see \S\ref{NR}).

\subsection{Vincenti \& Baldwin}

Vincenti \& Baldwin \cite{vb62} perform an analysis similar to that of Bogdan et al. \cite{bog96}, with the additional assumptions of negligible radiation pressure and time scales much greater than the time scale for coupling between the radiation and material (in our notation, $r \ll 1$ and $\tau_c \gg 1$). The correspondence between their notation and ours is
\be\label{VBC1}
ic_j \leftrightarrow \tau_a\tau_k^{-1},
\ee
\be
\nbo \leftrightarrow \frac{a}{rc}
\ee
and
\be\label{VBC3}
\nbu \leftrightarrow \tau_a.
\ee

We demonstrate in Appendix \ref{VB} that their dispersion relation is
\be\label{VB1}
\tau_a^{-2} - \tau_k^{-2} - i16r\tau_c\left(\gamma \tau_a^{-2} - \tau_k^{-2} \right) \int_0^1 d\mu \; \frac{\mu^2 }{\tau_k^2 + \mu^2} = 0,
\ee
which is equivalent to the transcendental relation
\be\label{VBDR}
\tau_a^{-2} - \tau_k^{-2} - i16r\tau_c(\gamma \tau_a^{-2} - \tau_k^{-2}) \left(1 -  \tau_k \tan^{-1}\tau_k^{-1}\right)  = 0.
\ee
This reduces to the dispersion relation of Bogdan et al. \cite{bog96} in the quasistatic limit (i.e., the determinant of the matrix at the top of p. 884 of \cite{bog96} with $s = i\omega$ and $\zeta_{\rm dop} = \zeta_{\rm dyn} = \zeta_{\rm tof} = 0$). Vincenti \& Baldwin \cite{vb62} replace the integral in expression (\ref{VB1}) with a single value of $\mu = 0.64$ and $d\mu = 0.813$ to obtain an approximate dispersion relation upon which they base the remainder of their analysis:
\be
\left(1 - i13r\tau_c\right)\tau_k^{-4} + \left(2.44 - \tau_a^{-2} + i13 \gamma r\tau_c\tau_a^{-2}\right)\tau_k^{-2} - 2.44\tau_a^{-2} = 0.
\ee
This dispersion relation captures the correct phase speed for the acoustic mode in regions a-d and the damping length to within $0.1\%$ (regions a and c), $19\%$ (region b) or $20\%$ (region d). For regions A and B, it is the damping length that is captured correctly with the phase speeds correct to within $0.1\%$. For regions E and F, the errors in the phase speed and damping length are $19\%$ and $27\%$, respectively. Regions e-g, C-D and G-H are not captured due to the additional simplifying assumptions made.

\subsection{Su \& Olson}

The problem considered by Su \& Olson \cite{so96} is also very similar to the one analyzed here. For late times they are essentially the same problem, although the analysis in Su \& Olson \cite{so96} is considerably more restrictive in its applicability due to additional simplifying assumptions (such as the neglect of hydrodynamic motions). After an initial boundary layer in time, their solutions have the form of a constant background plus a small perturbation (see their equation [31]).\footnote{This implies that their assumption of a cubic temperature dependence for the heat capacity is only necessary for the initial temporal boundary layer.} The correspondence between their notation and ours is
\be
\epsilon \leftrightarrow 16\gamma r,
\ee
\be
s \leftrightarrow \frac{i}{16\gamma r \tau_c},
\ee
and
\be
-\beta^2 \leftrightarrow \frac{1}{3} \tau_k^{-2},
\ee
Their equation (21) is a dispersion relation between wavenumber and frequency. In our notation it is
\be
(1 - i16\gamma r \tau_c) \tau_k^{-2} + 3(1 + i\tau_c^{-1} + 16\gamma r) = 0.
\ee
This captures the radiative diffusion modes in regions B, C, F and G. One can show that the velocity perturbation for these modes is usually much smaller than the perturbations in temperature and radiation energy, thus validating their neglect of hydrodynamic motions.\footnote{There are portions of region C, particularly for values of $\epsilon$ larger than they consider, in which hydrodynamic motions become important and this assumption breaks down.}

Perhaps the most important distinction between our approach and that taken by Su \& Olson \cite{so96} is that their $\omega$ is imaginary whereas ours is real; as a result, their modes do not propagate. With that caveat in mind, their inverse Laplace transform operation can be viewed (at late times) as a linear superposition of radiative diffusion modes. This superposition introduces two complications: 1) the high frequency components make numerical evaluation of the semi-analytical result difficult, and 2) there is an additional source of error for a code comparison due to the attempt to represent a continuum of frequencies with a discretization. Both of these complications are introduced without testing the code beyond what one can do with a single mode. An advantage of the approach taken by Su \& Olson \cite{so96}  is that it allows for a wider range of initial conditions since it captures the temporal boundary layer. 

\section{Numerical Results}
\label{NR}

The numerical implementation of the solutions described in the previous section requires driving the boundary of a one-dimensional computational domain at frequency $\omega$, with material and radiation fluids satisfying the assumptions described in \S\ref{AE}. We employ the code Kull for our numerical calculations, a description of which can be found in reference \cite{kull00}. Kull is an Arbitrary Lagrange Eulerian (ALE) code, although we only present results with Kull in Lagrangian mode. We drive both the radiation temperature and the material velocity at one boundary, and we use a Milne boundary condition on the radiation \cite{cas04} at the opposite boundary. Kull does not support an outflow boundary condition for the material, so we simply fix the velocity at the opposite boundary and stop the calculation before the perturbation reaches the far end of the grid.\footnote{Driving at one end with outflow at the other end would be the appropriate boundary conditions for an Eulerian calculation as well.} 

The dimensionless measures of temperature and density for a system in LTE are $a/c$ and $r$. For $\gamma = 5/3$ and a mean molecular mass of $0.6$, the corresponding physical temperature and density scales are
\be
T = 4 \times 10^{12} \left(\frac{a}{c}\right)^2 \; \K
\ee
and
\be
\rho = 6 \times 10^{-24} \left(\frac{T}{1K}\right)^3 r^{-1} \; \gm \cm^{-3} = 3 \times 10^{14} \left(\frac{a}{c}\right)^6 r^{-1} \; \gm \cm^{-3}.
\ee
We conduct nearly all of our runs with $a = 10^{-4}c$, corresponding to a mean temperature of $4\times10^4\K$.

We define our computational domain to be a fixed fraction or multiple of the perturbation wave length $\lambda$. For a given solution to the dispersion relation (\ref{DR}), the opacity is given by 
\be
\chi = \frac{2\pi}{\lambda \, {\rm Re}[\tau_k^{-1}]}
\ee
and the driving frequency by
\be
\omega = \frac{2\pi a}{\lambda \tau_a \, {\rm Re}[\tau_k^{-1}]}.
\ee
The numerical length scale can be associated with a physical length scale by calculating $\chi$ based upon a particular frequency-integrated opacity. The physical time scale is then determined by this length scale and the speed of light.

Results for the radiative acoustic wave are shown in Figures~\ref{fa}-\ref{ff} (regions a-f of Figure~\ref{fra}). The points are the numerical solutions and the solid lines are the analytical solutions. The boundary at which the driving is applied is on the left, the computational domain is ten wave lengths, and we run the simulation for ten wave periods to ensure that reflection off the right boundary does not influence our results. We plot the density perturbation at the end of each run. All of these results are at a resolution of $80$ zones per wave length. The discrepancies at the right hand side of these figures are due to initial transients that are not captured by the analytical solution.\footnote{The use of outflow boundary conditions would allow these transient features to propagate out of the computational domain.}

Results for the radiative diffusion wave are shown in Figures~\ref{fA}-\ref{fF} (regions A-F of Figure~\ref{frd}). The computational domain is ten wave lengths in regions A, B and D, although we only plot the first wave length since that is the length over which the perturbation is damped. We integrate these runs for ten wave periods and the number of zones per wave length is again $80$. Since the damping length in regions E and F is much smaller than a wave length (by a factor of $\sim 10^{-2}$), we use a computational domain of one half of a wave length for these runs, to give $\sim 16$ zones per damping length. For all of the radiation diffusion runs we plot the radiation temperature perturbation at the end of each run.

We include a high resolution result for region C (Figure~\ref{fC}) to demonstrate the significant numerical cost that can be required to obtain an accurate result. The computational domain for this run is one wave length and we integrate for two wave periods. The number of zones per wave length is $1600$ ($20$ times our nominal value), and the time step was set to the diffusion time scale, as opposed to the implicit time integration that was employed for the other runs.

Figure~\ref{fE} highlights the fact that coupling to both modes can occur, making it difficult to isolate a single mode. We demonstrate this explicitly in Figure~\ref{fEcoupling}, which shows Kull results at various resolutions along with the analytical solution for both the acoustic mode and the diffusion mode. While we are attempting to drive the diffusion mode, it is clear that the acoustic mode is being excited as well, with the amplitude of the excitations decreasing as we increase the resolution.\footnote{The use of outflow boundary conditions on the material would likely reduce this effect considerably.}

Another consideration when comparing the numerical results to the analytical solution is that different perturbations dominate in different regions of parameter space, and comparisons to the analytical solution are more robust when based upon the dominant perturbation. The density and velocity perturbations generally dominate for the acoustic mode, and the radiation temperature perturbation for the diffusion mode. Another implication of this is that care must be taken when setting the overall amplitude of the perturbations; a small amplitude for the hydrodynamic variables may translate into a nonlinear amplitude for the radiation energy density, or vice versa.

Figures~\ref{fgs1}-\ref{fHs2} demonstrate that flux-limited diffusion does not preserve causality for small amplitude waves. Figures~\ref{fgs1} and \ref{fgs2} are results for an acoustic wave (region g of Figure~\ref{fra}); these were run at $a = 0.1 c$ to make this region of parameter space more computationally accessible. For a sufficiently small amplitude (Figure~\ref{fgs1}), the numerical solution converges to the analytical solution under the diffusion approximation, both with and without a flux limiter. A driven wave is superluminal if its wavelength exceeds the wavelength of free-streaming radiation at the same frequency; i.e., $v_p > c$ for
\be
k < \frac{\omega}{c}.
\ee
The dashed line in Figure~\ref{fgs1} shows a wave with a phase velocity equal to $c$, demonstrating the superluminal nature of the excited wave. As the amplitude is increased (Figure~\ref{fgs2}), the wave begins to steepen into a shock, with the flux limiter doing nothing to limit the amplitude of the radiation temperature perturbation.

Figures~\ref{fHs1} and \ref{fHs2} show similar results for a diffusion wave (region H of Figure~\ref{frd}). Figure~\ref{fHs1} again demonstrates that the numerical results (with and without a flux limiter) are converging to the analytical solution under the diffusion approximation, which is superluminal in this region of parameter space (albeit damped over a wavelength). Figure~\ref{fHs2} indicates that as the amplitude of the wave approaches the nonlinear regime in this case, the flux limiter begins to shorten the wavelength. 

\section{Summary and Discussion}
\label{SD}

We have conducted a systematic comparison of a perturbation analysis of the equations of radiation hydrodynamics with the Lagrangian code Kull in a wide range of parameter space. We have demonstrated that these solutions are a useful benchmark for testing any radiation hydrodynamics code. The most important issues to keep in mind when conducting such a test are 1) flux-limited diffusion does not capture these solutions in the free-streaming limit, 2) coupling to both modes can occur, making comparison with a single mode somewhat difficult and 3) care must be taken in setting the perturbation amplitude, since either the radiation or hydrodynamic perturbations can dominate the others by orders of magnitude.

\begin{figure}
\psfrag{y}[][][1.5]{$\frac{\delta \rho}{\rho_0}$}
\psfrag{x}[][][1.5]{$\frac{x}{\lambda}$}
\centering
\includegraphics[width=6.0in]{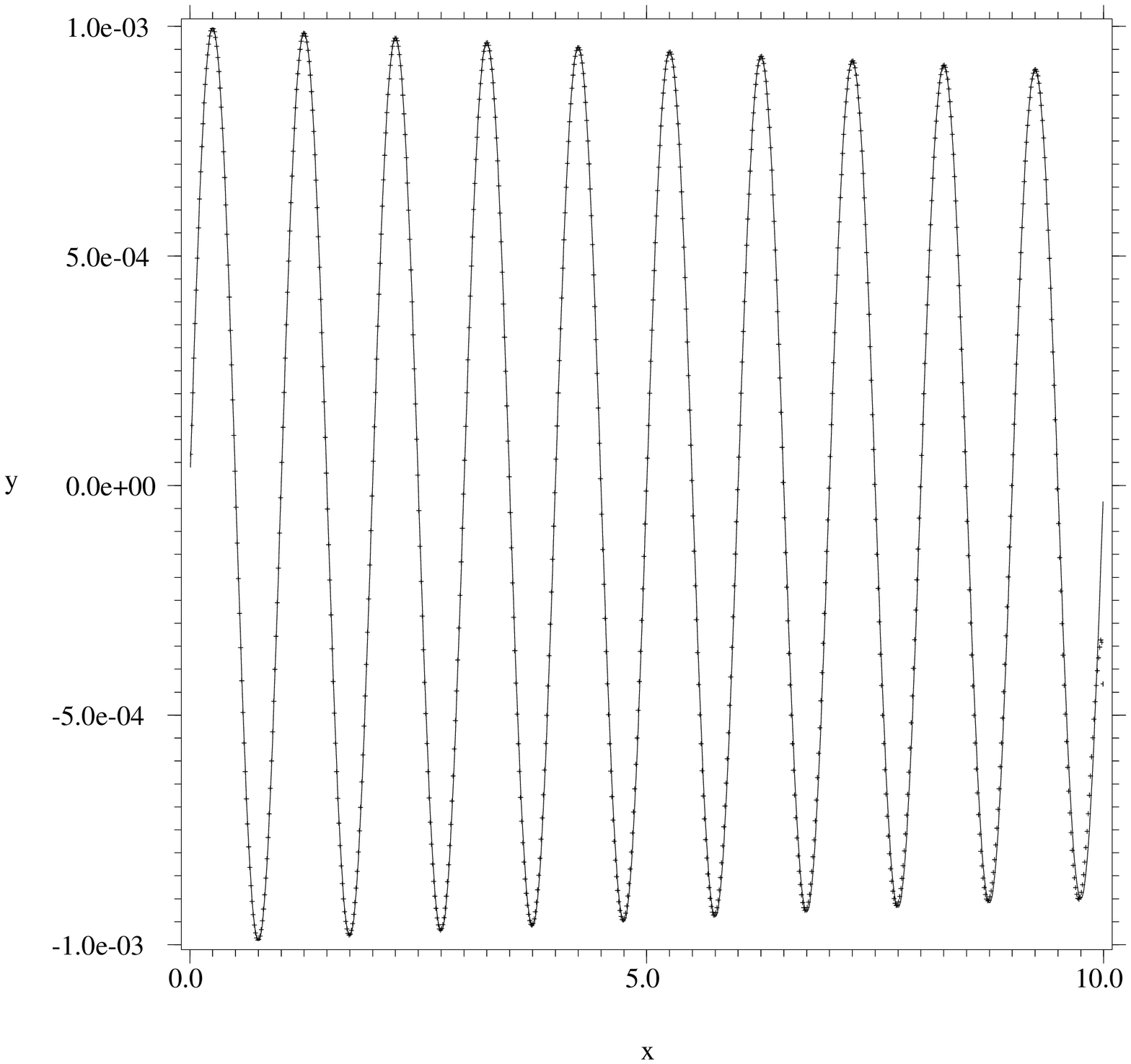}
\caption{Density perturbation in region a ($r = 10^{-3}$, $\tau_a = 10^4$) after ten wave periods. The points are the Kull results, and the solid line is the analytical solution.}
\label{fa}
\end{figure}

\begin{figure}
\psfrag{y}[][][1.5]{$\frac{\delta \rho}{\rho_0}$}
\psfrag{x}[][][1.5]{$\frac{x}{\lambda}$}
\centering
\includegraphics[width=6.0in]{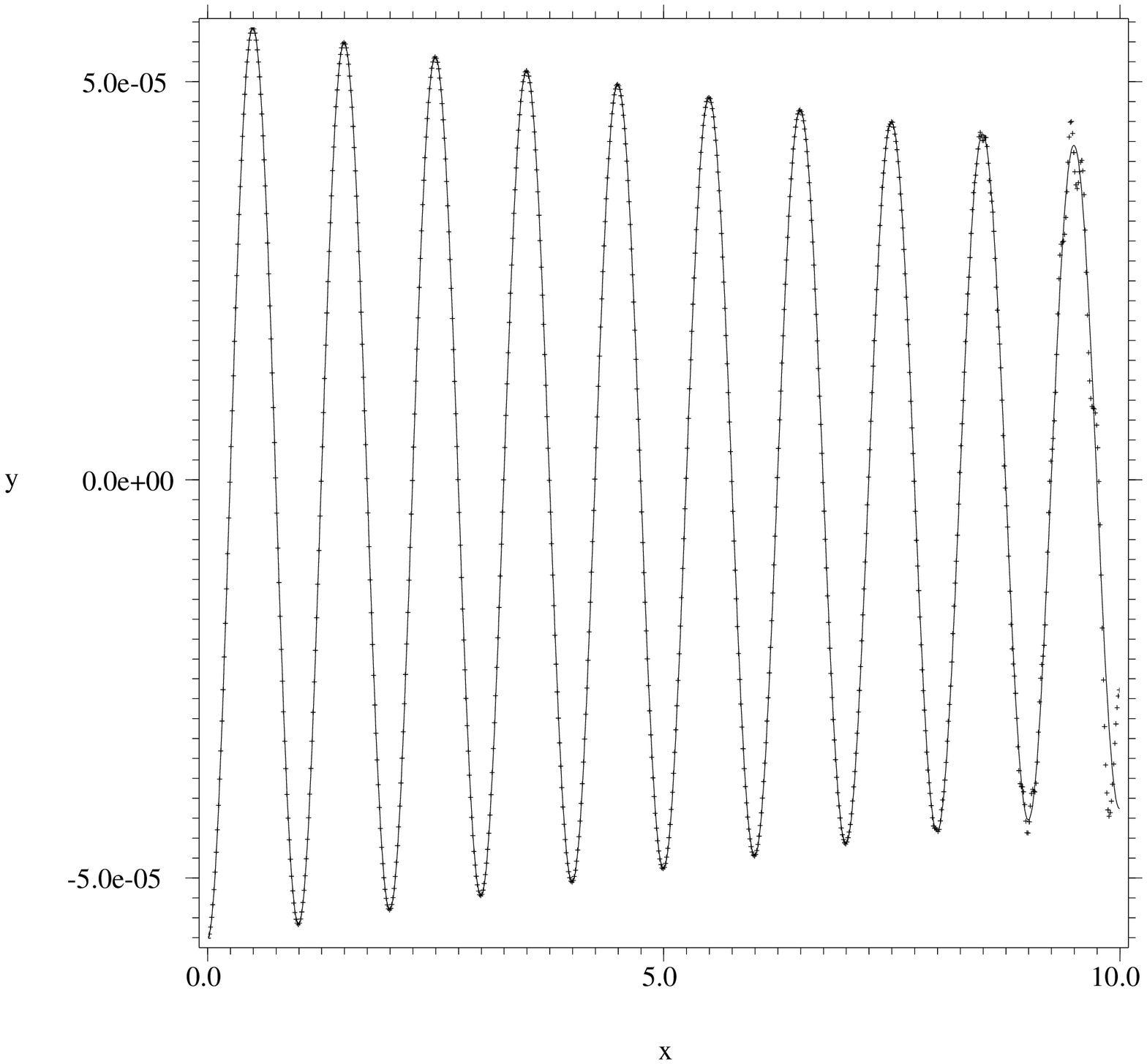}
\caption{Density perturbation in region b ($r = 10^{-5}$, $\tau_a = 10^{-2}$) after ten wave periods. The points are the Kull results, and the solid line is the analytical solution.}
\label{fb}
\end{figure}

\begin{figure}
\psfrag{y}[][][1.5]{$\frac{\delta \rho}{\rho_0}$}
\psfrag{x}[][][1.5]{$\frac{x}{\lambda}$}
\centering
\includegraphics[width=6.0in]{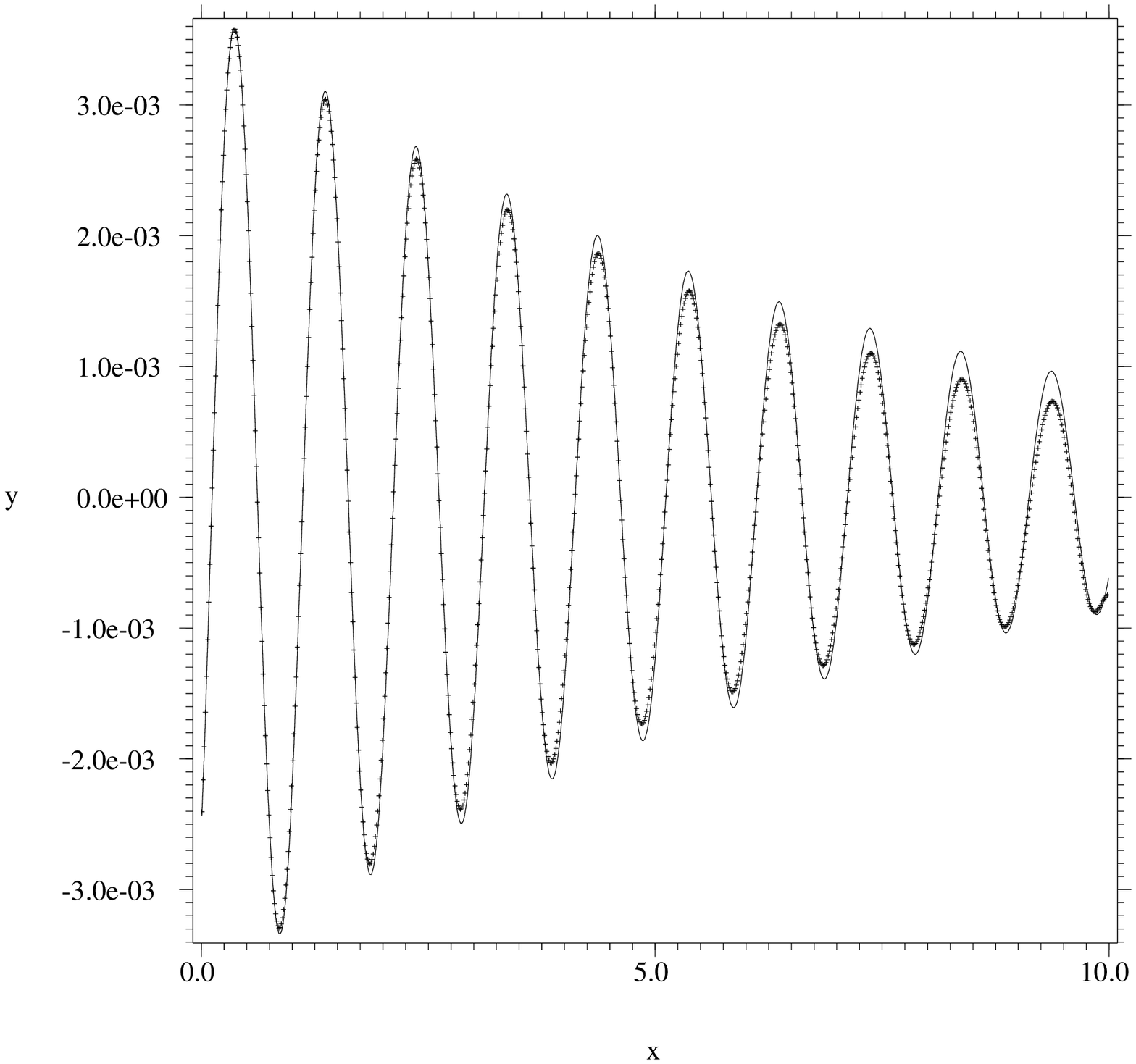}
\caption{Density perturbation in region c ($r = 10^{-3}$, $\tau_a = 10$) after ten wave periods. The points are the Kull results, and the solid line is the analytical solution.}
\label{fc}
\end{figure}

\begin{figure}
\psfrag{y}[][][1.5]{$\frac{\delta \rho}{\rho_0}$}
\psfrag{x}[][][1.5]{$\frac{x}{\lambda}$}
\centering
\includegraphics[width=6.0in]{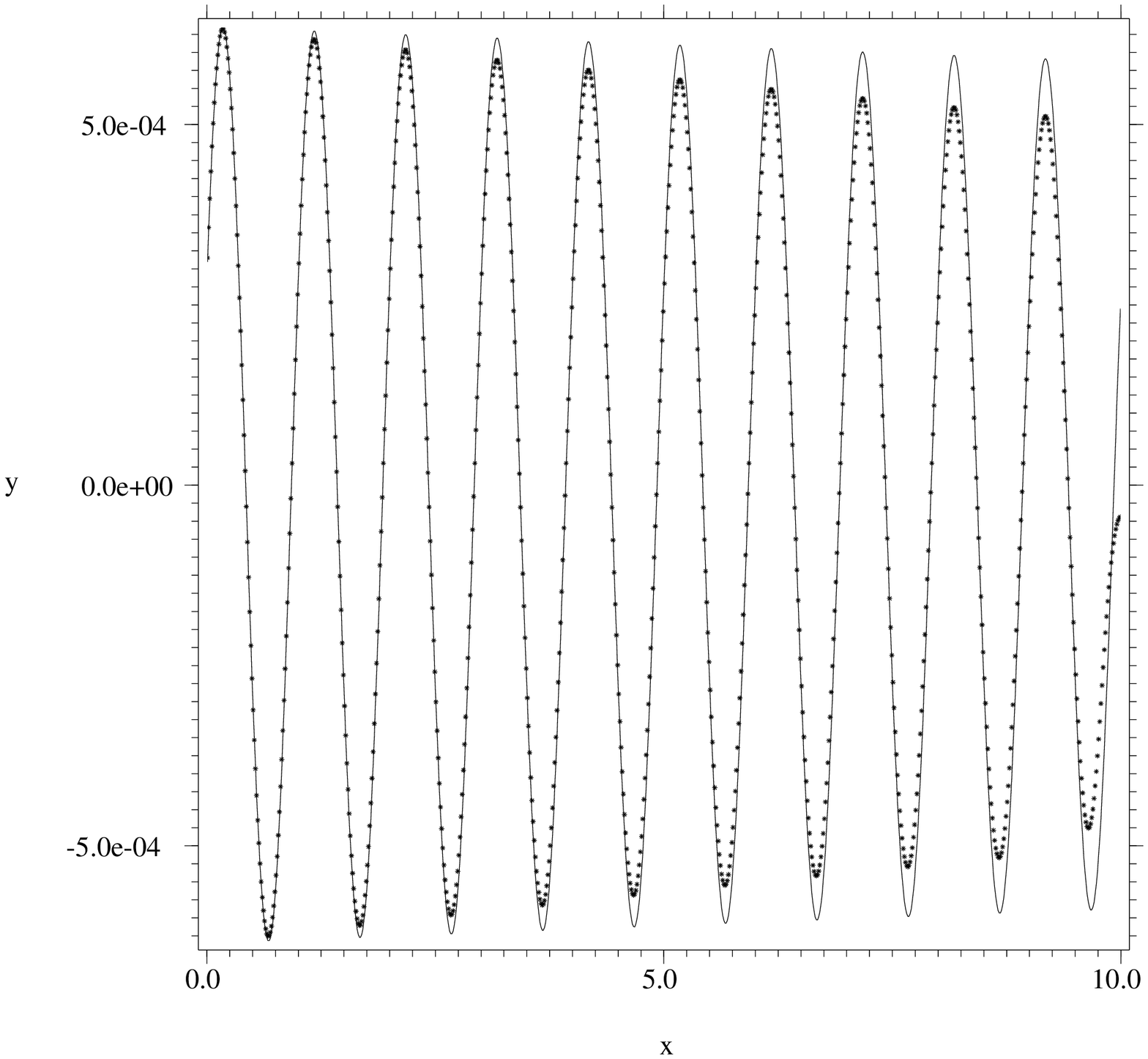}
\caption{Density perturbation in region d ($r = 10^{-1}$, $\tau_a = 10^{-2}$) after ten wave periods. The points are the Kull results, and the solid line is the analytical solution.}
\label{fd}
\end{figure}

\begin{figure}
\psfrag{y}[][][1.5]{$\frac{\delta \rho}{\rho_0}$}
\psfrag{x}[][][1.5]{$\frac{x}{\lambda}$}
\centering
\includegraphics[width=6.0in]{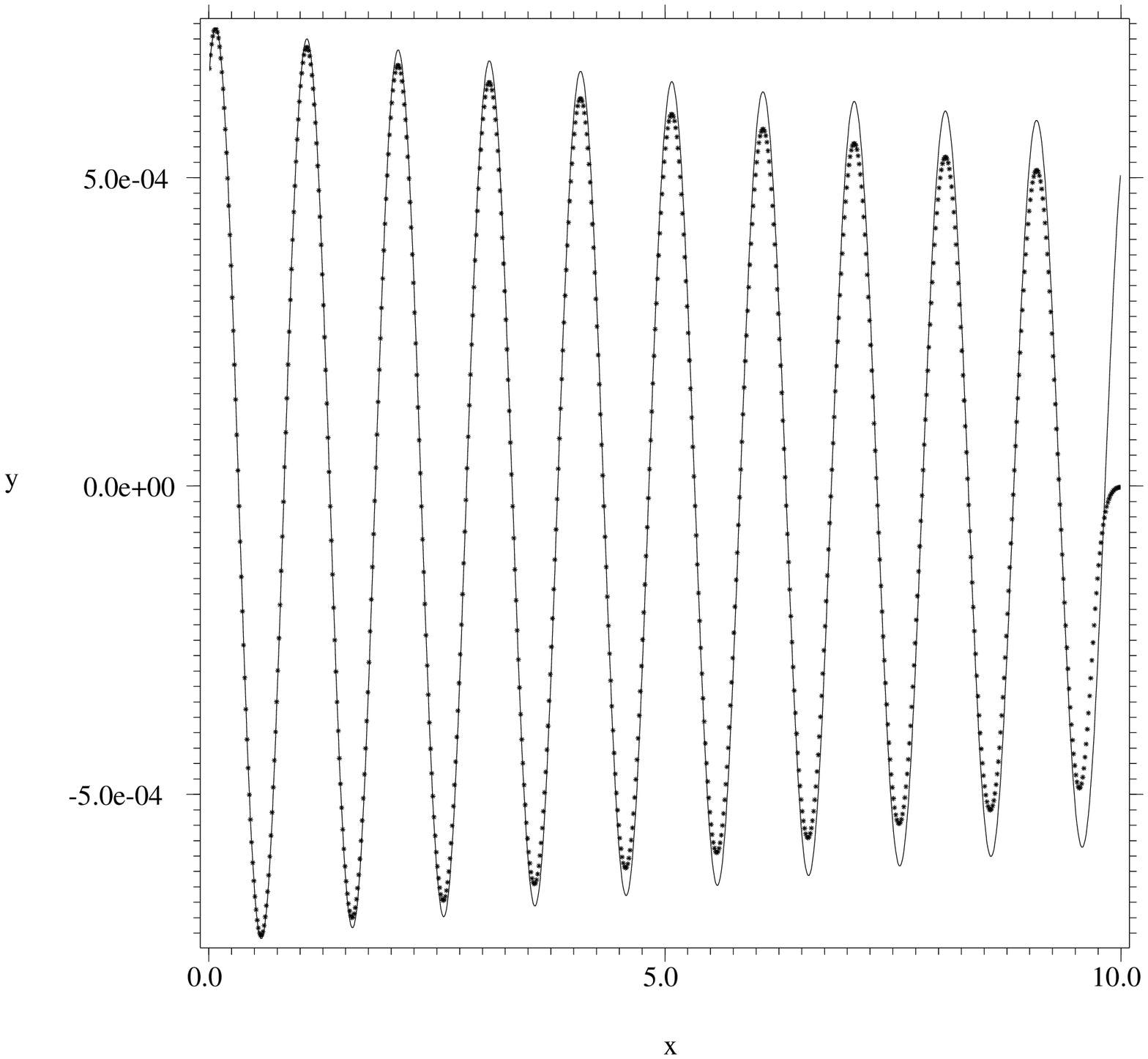}
\caption{Density perturbation in region e ($r = 10^3$, $\tau_a = 10^{-2}$) after ten wave periods. The points are the Kull results, and the solid line is the analytical solution.}
\label{fe}
\end{figure}

\begin{figure}
\psfrag{y}[][][1.5]{$\frac{\delta \rho}{\rho_0}$}
\psfrag{x}[][][1.5]{$\frac{x}{\lambda}$}
\centering
\includegraphics[width=6.0in]{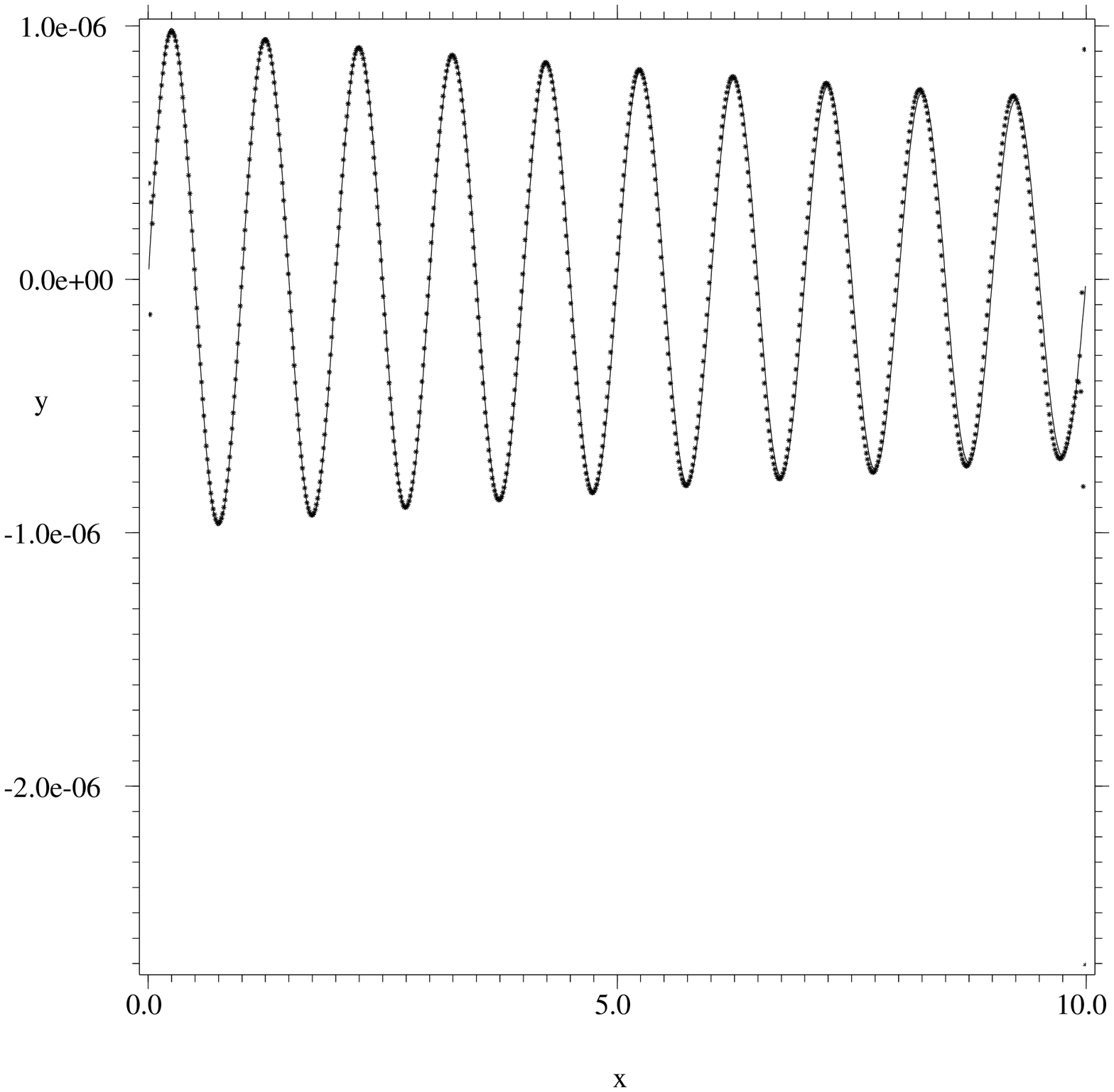}
\caption{Density perturbation in region f ($r = 10$, $\tau_a = 10^4$) after ten wave periods. The points are the Kull results, and the solid line is the analytical solution.}
\label{ff}
\end{figure}

\begin{figure}
\psfrag{y}[][][1.5]{$\frac{\delta T_r}{T_0}$}
\psfrag{x}[][][1.5]{$\frac{x}{\lambda}$}
\centering
\includegraphics[width=6.0in]{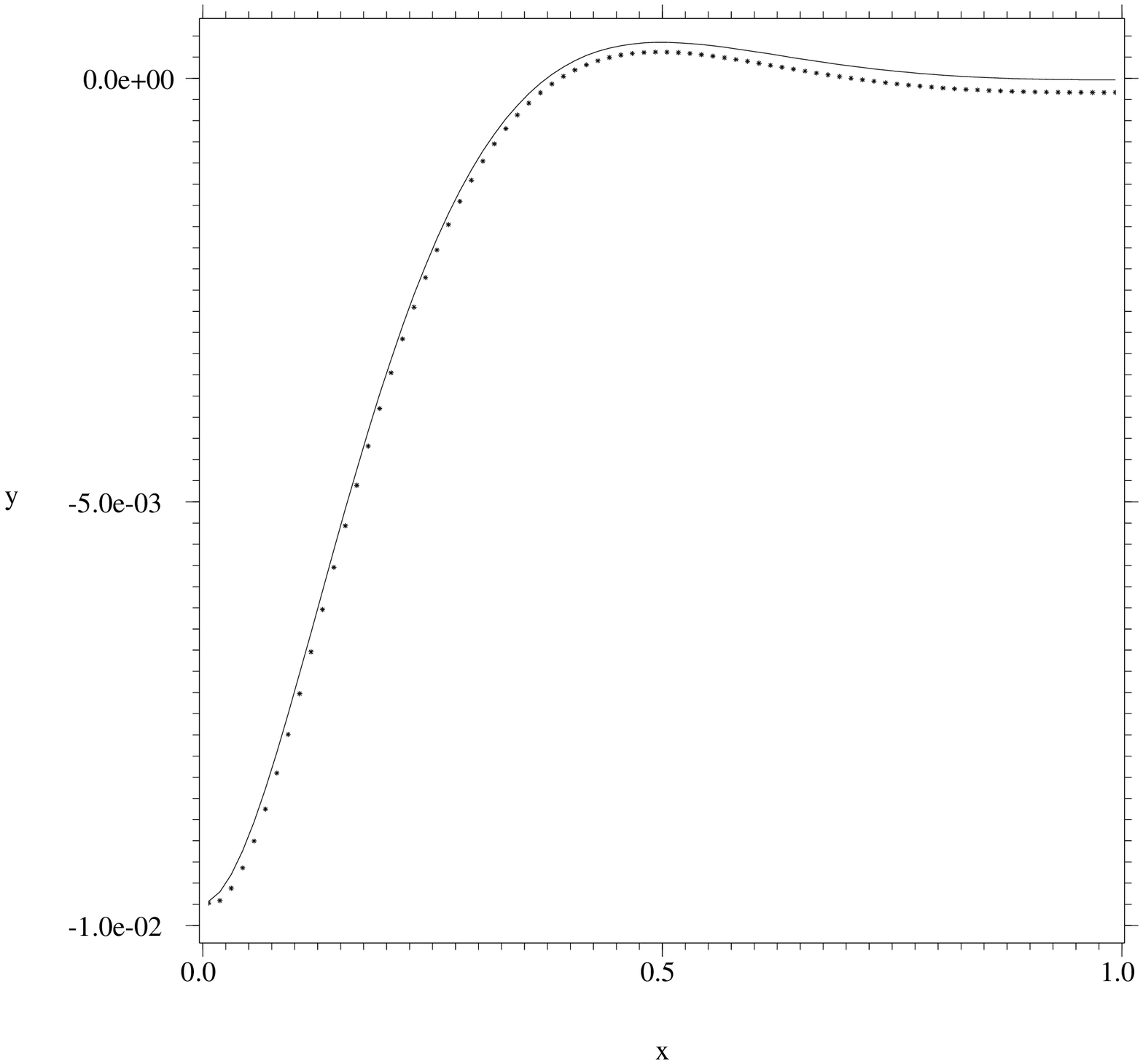}
\caption{Radiation temperature perturbation in region A ($r = 10^{-3}$, $\tau_a = 10^4$) after ten wave periods. The points are the Kull results, and the solid line is the analytical solution. Only one-tenth of the computational domain is shown.}
\label{fA}
\end{figure}

\begin{figure}
\psfrag{y}[][][1.5]{$\frac{\delta T_r}{T_0}$}
\psfrag{x}[][][1.5]{$\frac{x}{\lambda}$}
\centering
\includegraphics[width=6.0in]{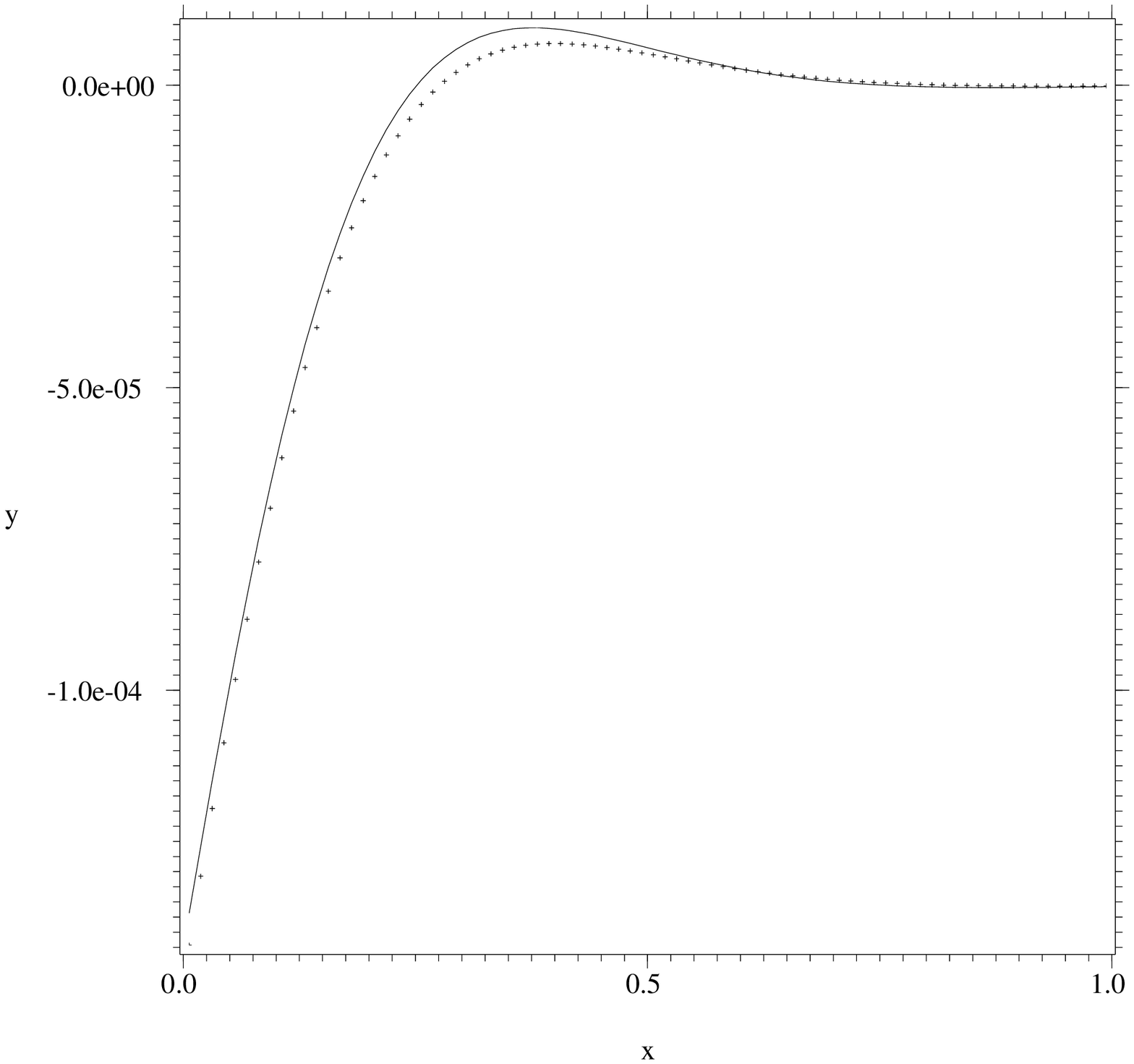}
\caption{Radiation temperature perturbation in region B ($r = 10^{-3}$, $\tau_a = 1$) after ten wave periods. The points are the Kull results, and the solid line is the analytical solution. Only one-tenth of the computational domain is shown.}
\label{fB}
\end{figure}

\begin{figure}
\psfrag{y}[][][1.5]{$\frac{\delta T_r}{T_0}$}
\psfrag{x}[][][1.5]{$\frac{x}{\lambda}$}
\centering
\includegraphics[width=6.0in]{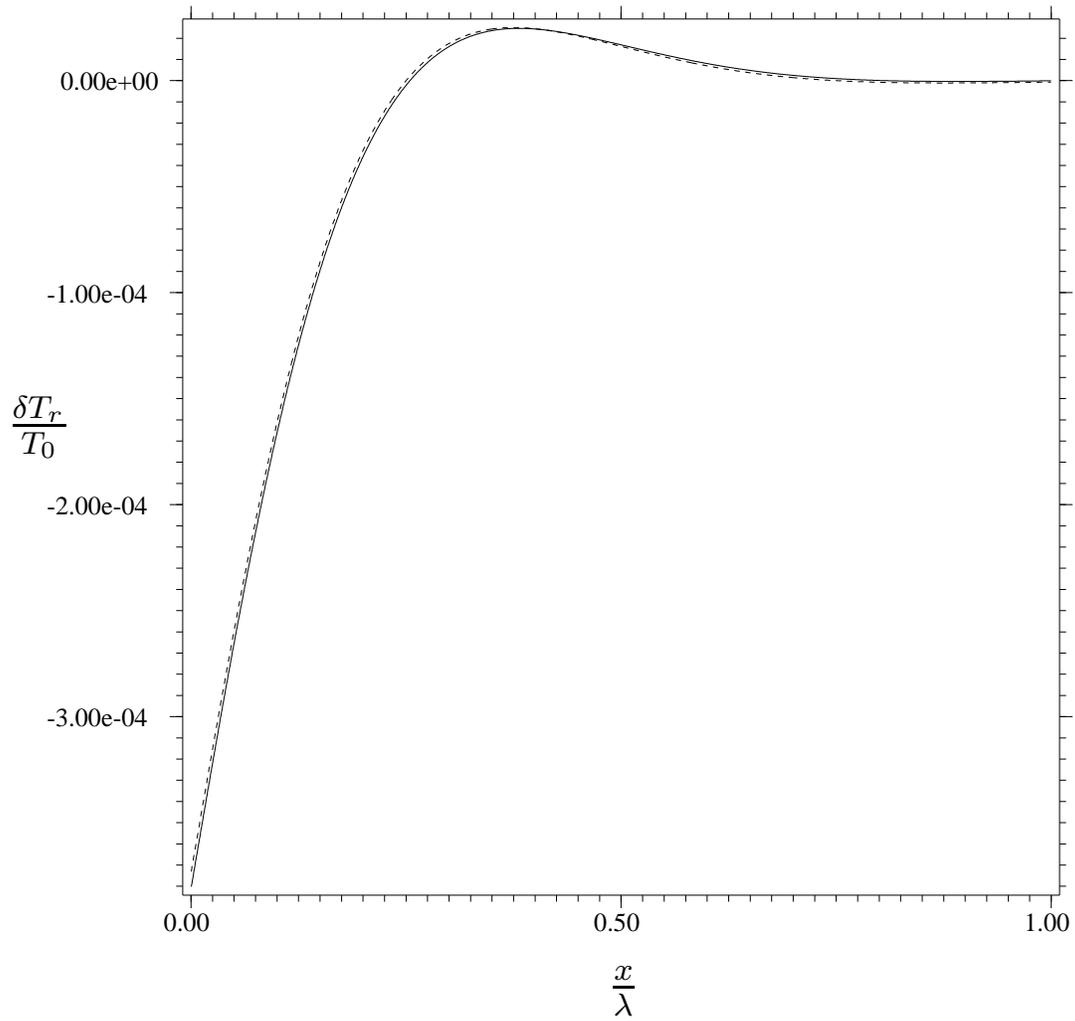}
\caption{Radiation temperature perturbation in region C ($r = 10^{-3}$, $\tau_a = 10^4$) after two wave periods. The dotted line is the Kull result, and the solid line is the analytical solution.}
\label{fC}
\end{figure}

\begin{figure}
\psfrag{y}[][][1.5]{$\frac{\delta T_r}{T_0}$}
\psfrag{x}[][][1.5]{$\frac{x}{\lambda}$}
\centering
\includegraphics[width=6.0in]{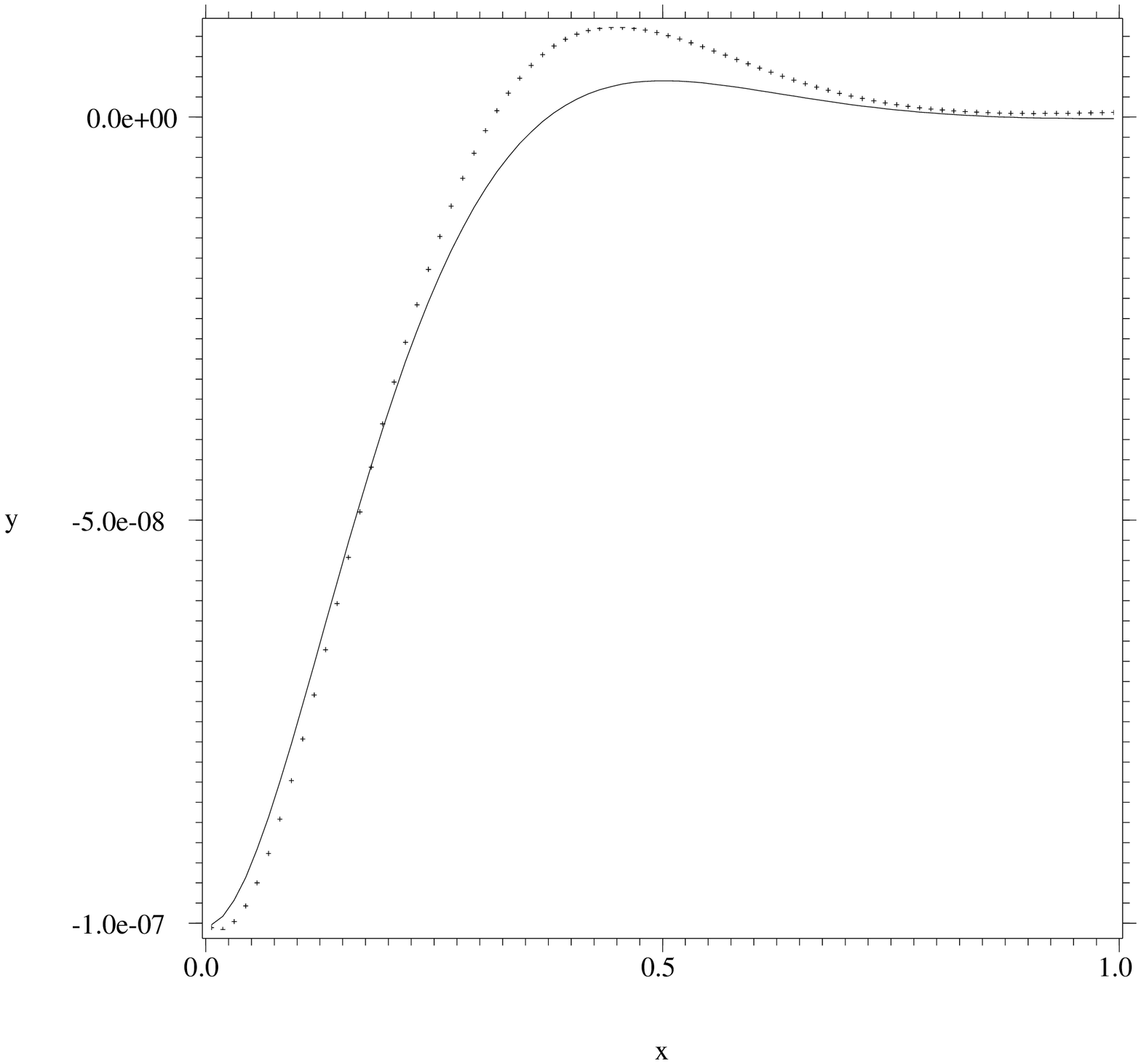}
\caption{Radiation temperature perturbation in region D ($r = 10^2$, $\tau_a = 10^3$) after ten wave periods. The points are the Kull results, and the solid line is the analytical solution. Only one-tenth of the computational domain is shown.}
\label{fD}
\end{figure}

\begin{figure}
\psfrag{y}[][][1.5]{$\frac{\delta T_r}{T_0}$}
\psfrag{x}[][][1.5]{$\frac{x}{\lambda}$}
\centering
\includegraphics[width=6.0in]{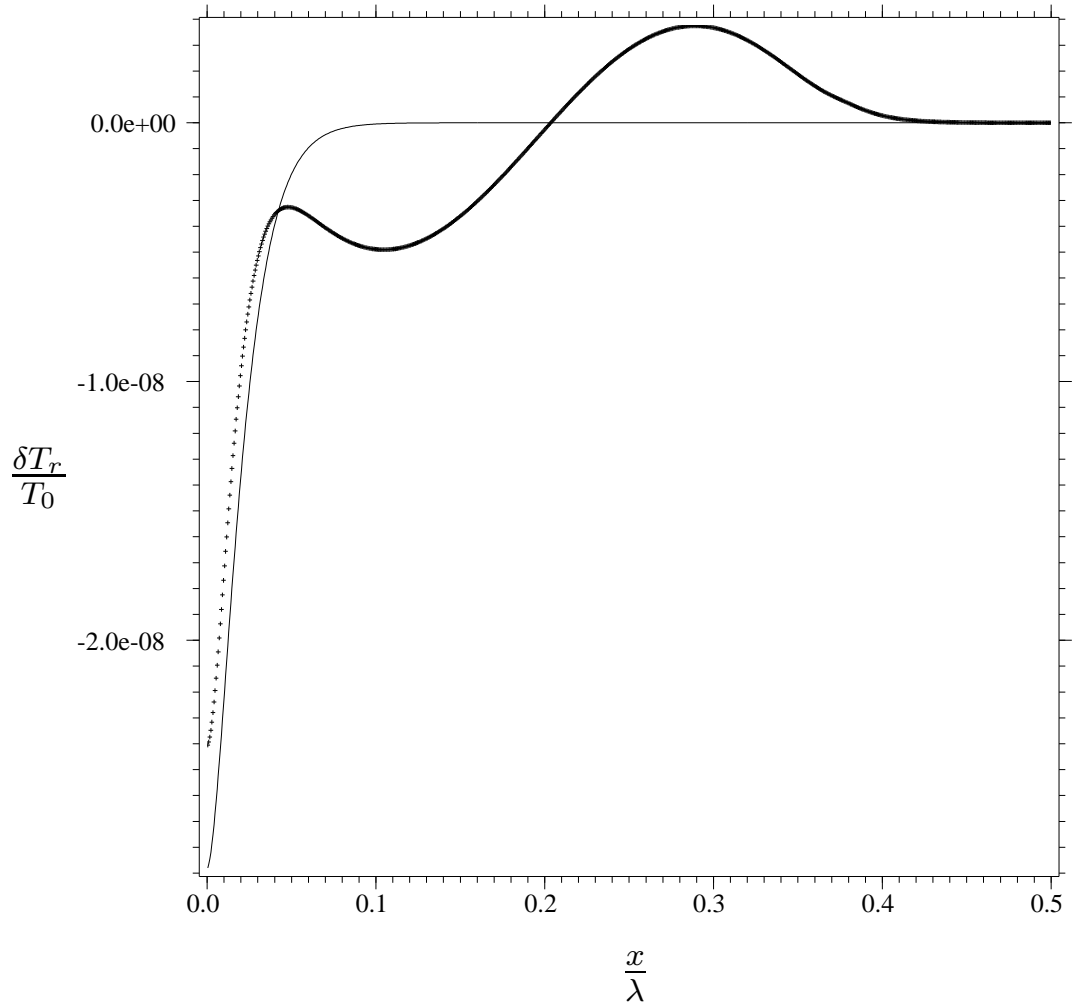}
\caption{Radiation temperature perturbation in region E ($r = 3 \times 10^{-7}$, $\tau_a = 3$) after one wave period. The points are the Kull results, and the solid line is the analytical solution.}
\label{fE}
\end{figure}

\begin{figure}
\psfrag{y}[][][1.5]{$\frac{\delta T_r}{T_0}$}
\psfrag{x}[][][1.5]{$\frac{x}{\lambda}$}
\centering
\includegraphics[width=6.0in]{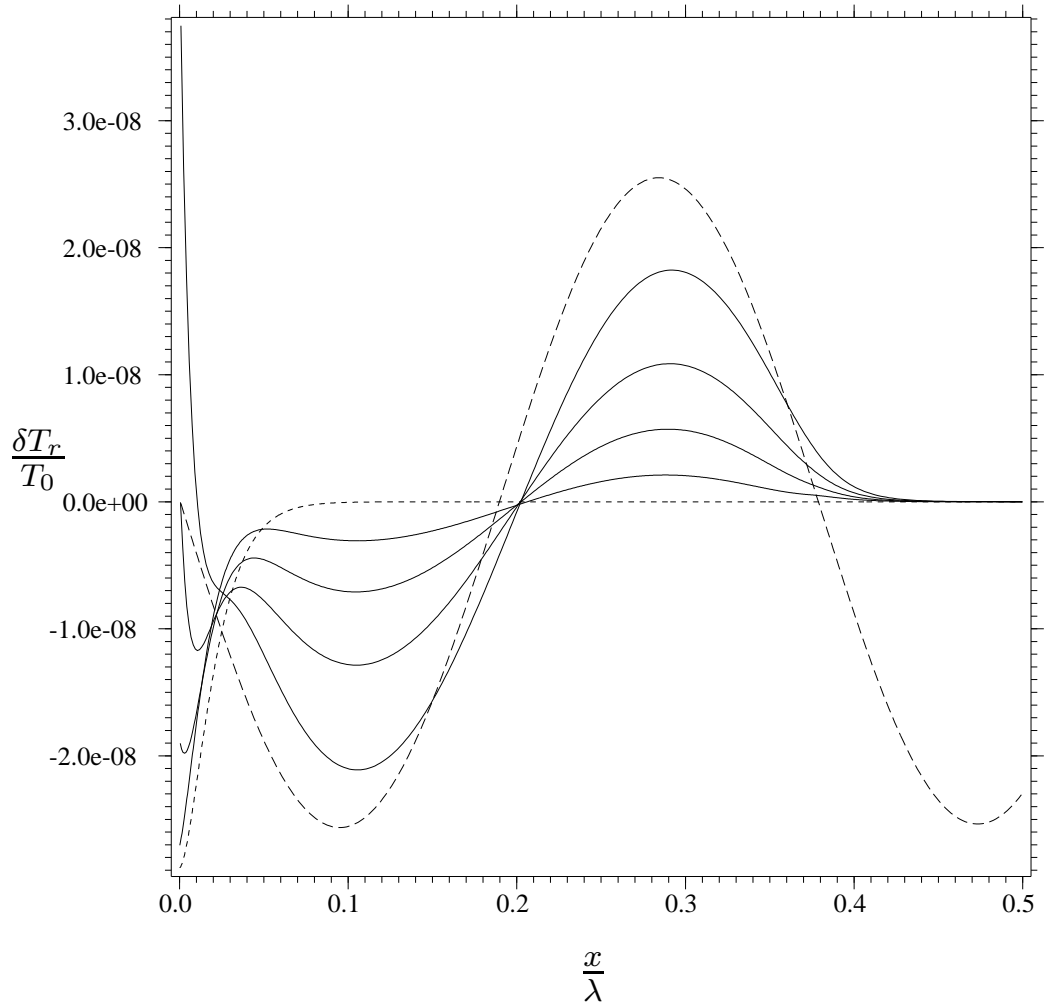}
\caption{Radiation temperature perturbation in region E after ten wave periods. The solid lines are Kull results with resolution increasing from top to bottom, the dotted line is the analytical solution for the diffusion mode, and the dashed line is the analytical solution for the acoustic mode.}
\label{fEcoupling}
\end{figure}

\begin{figure}
\psfrag{y}[][][1.5]{$\frac{\delta T_r}{T_0}$}
\psfrag{x}[][][1.5]{$\frac{x}{\lambda}$}
\centering
\includegraphics[width=6.0in]{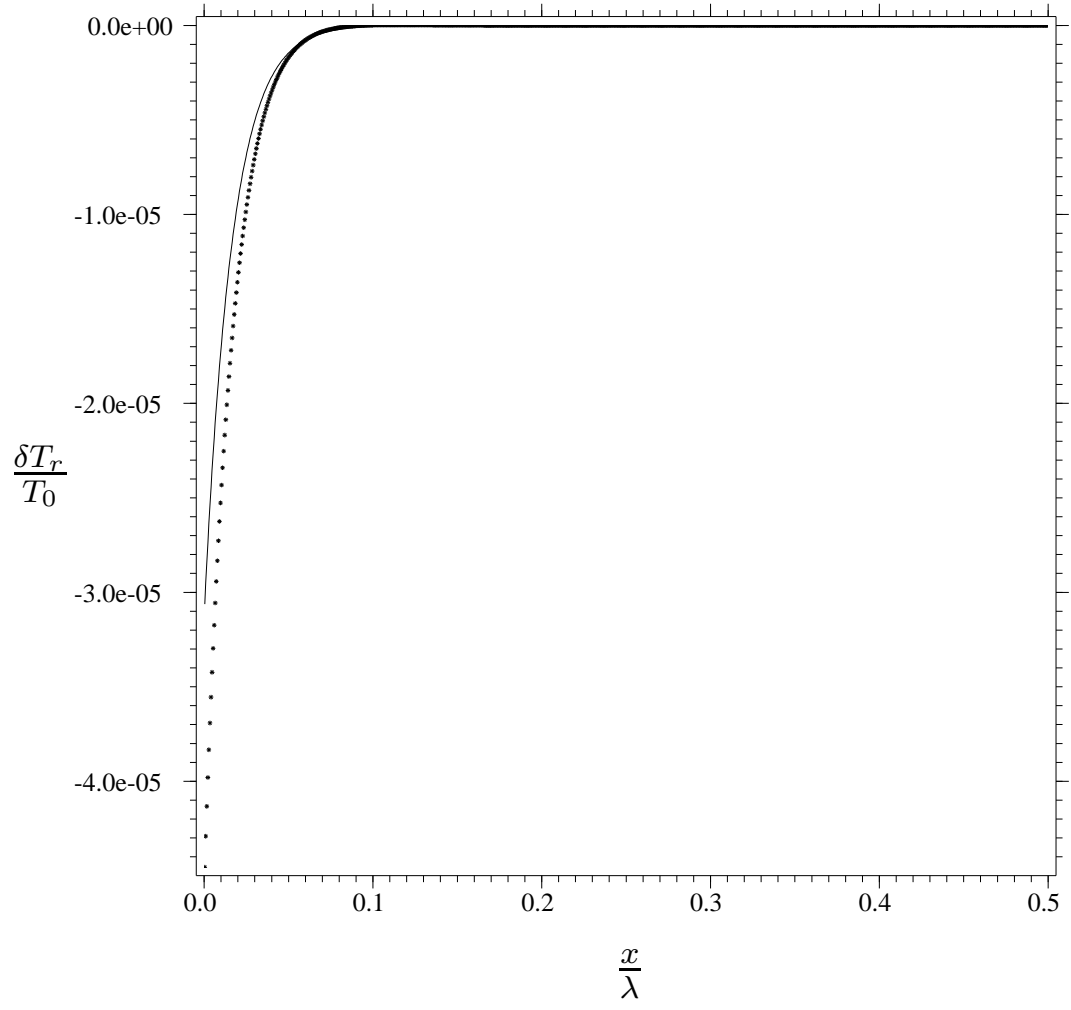}
\caption{Radiation temperature perturbation in region F ($r = 3 \times 10^{-6}$, $\tau_a = 3 \times 10^{-1}$) after one half of a wave period. The points are the Kull results, and the solid line is the analytical solution.}
\label{fF}
\end{figure}

The primary purpose of a numerical test like the one we have studied is to investigate the convergence properties of a numerical discretization. Since our focus has been on the test itself rather than on the convergence properties of Kull, and due to the complications of transient effects and mode coupling, we have not included any convergence results here. We have investigated the convergence properties of Kull for most of the solutions shown in Figures~\ref{fa}-\ref{fF} by calculating the L2 norm of the error between the numerical and analytical results (excluding the final wavelength to minimize transient effects). Kull converges at second order in most regions of parameter space, as expected, particularly when the hydrodynamics and radiation are weakly coupled. In regions of parameter space where the coupling between modes is strong, the convergence is weaker. A formal convergence test in Kull using the solutions described here would require the implementation of outflow boundary conditions.

\begin{figure}
\psfrag{y}[][][1.5]{$\frac{\delta T_r}{T_0}$}
\psfrag{x}[][][1.5]{$\frac{x}{\lambda}$}
\centering
\includegraphics[width=6.0in]{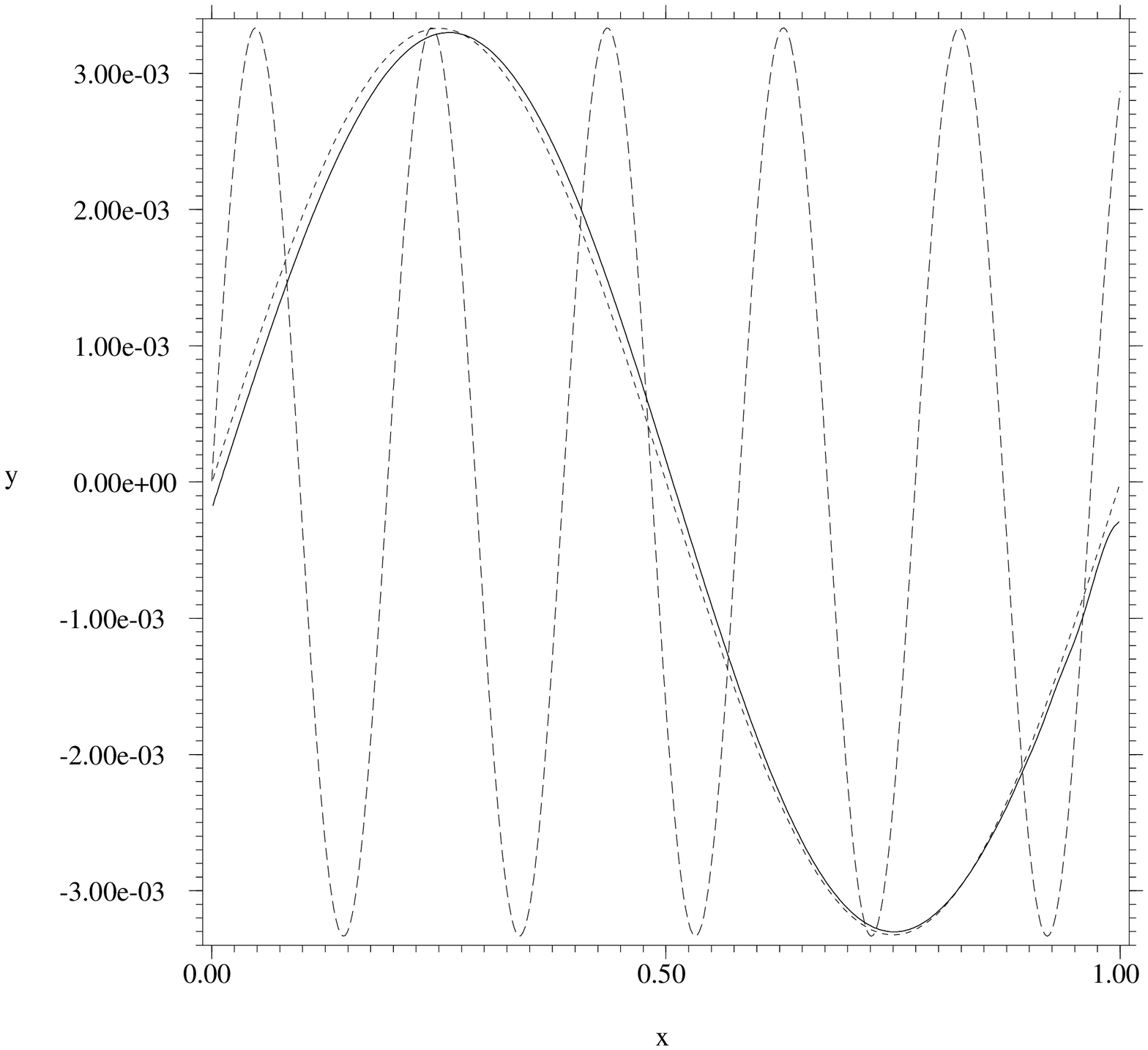}
\caption{Radiation temperature perturbation in region g ($r = 10^3$, $\tau_a = 1, a = 0.1 c$) after one wave period. The solid lines are the Kull results (both with and without a flux limiter; differences are $O[10^{-11}]$), and the dotted line is the analytical solution under the diffusion approximation. For reference purposes, the dashed line shows a wave at the same driving frequency with a phase velocity equal to the speed of light.}
\label{fgs1}
\end{figure}

\begin{figure}
\psfrag{y}[][][1.5]{$\frac{\delta T_r}{T_0}$}
\psfrag{x}[][][1.5]{$\frac{x}{\lambda}$}
\centering
\includegraphics[width=6.0in]{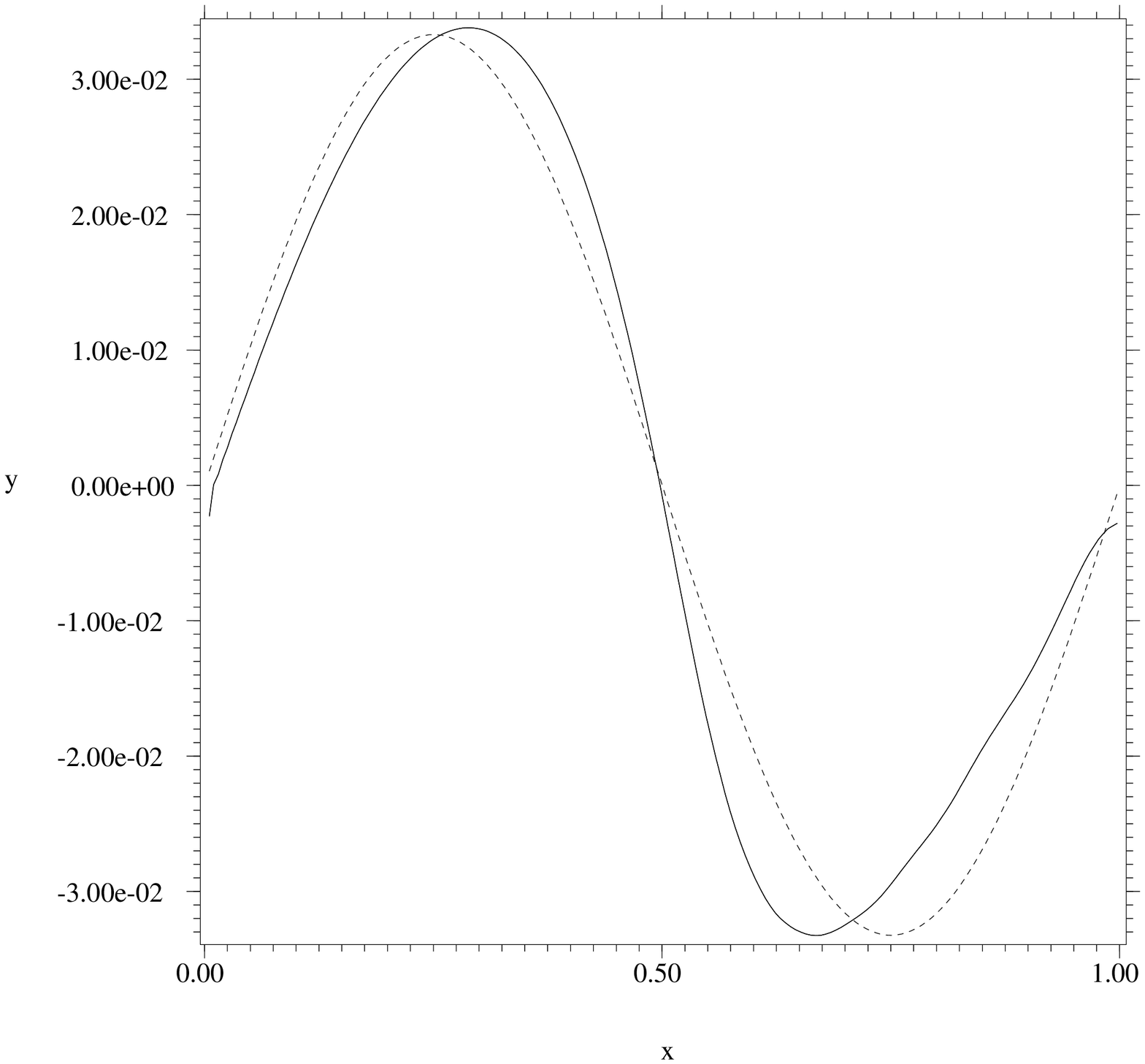}
\caption{Same as Figure~\ref{fgs1} with a higher perturbation amplitude. The differences between the numerical results with and without a flux limiter are $O(10^{-8})$.}
\label{fgs2}
\end{figure}

\begin{figure}
\psfrag{y}[][][1.5]{$\frac{\delta T_r}{T_0}$}
\psfrag{x}[][][1.5]{$\frac{x}{\lambda}$}
\centering
\includegraphics[width=6.0in]{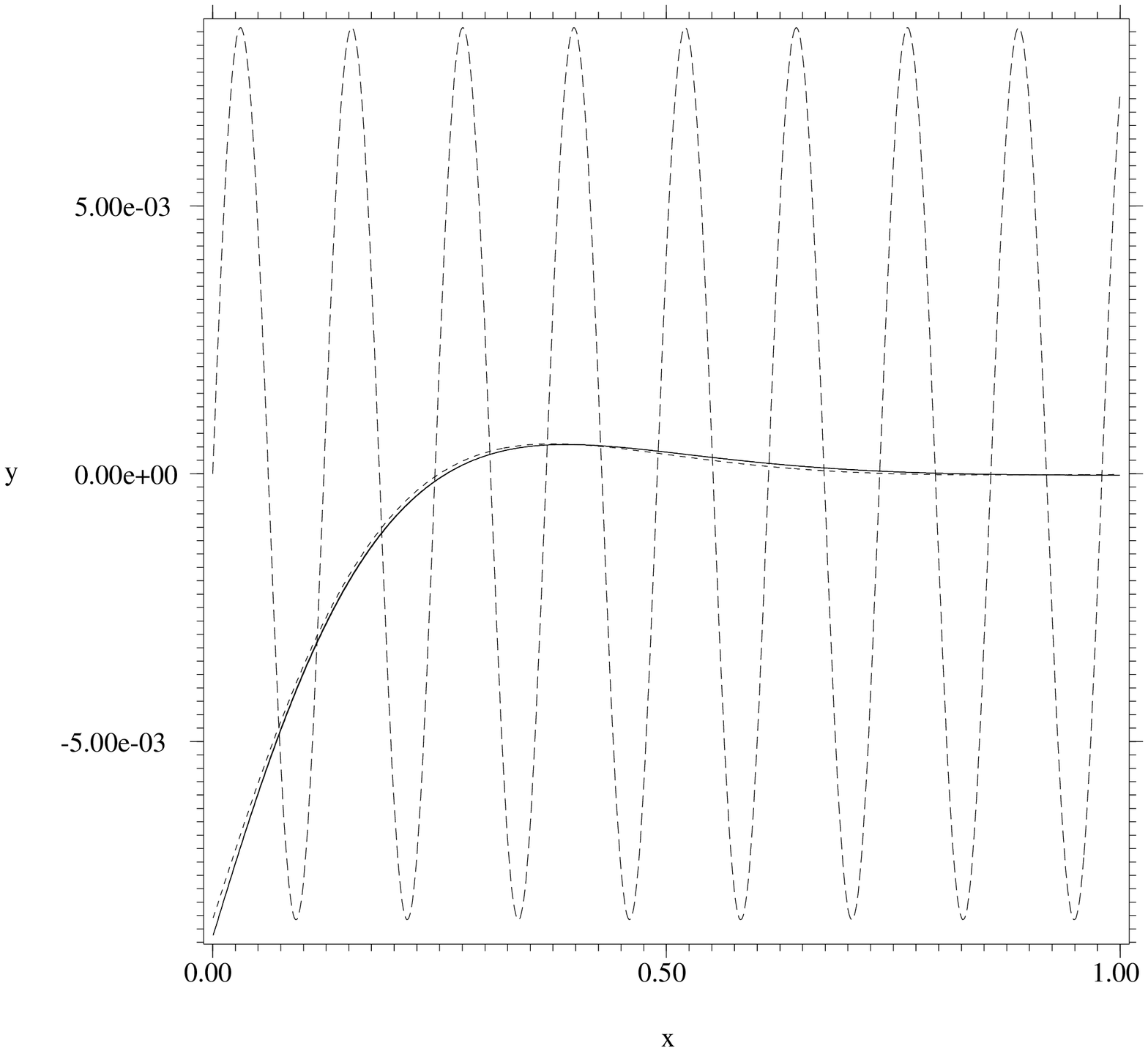}
\caption{Radiation temperature perturbation in region H ($r = 10^2$, $\tau_a = 10^{-6}$) after one wave period. The solid lines are the Kull results (both with and without a flux limiter; differences are $O[10^{-5}]$), and the dotted line is the analytical solution under the diffusion approximation. For reference purposes, the dashed line shows a wave at the same driving frequency with a phase velocity equal to the speed of light.}
\label{fHs1}
\end{figure}

\begin{figure}
\psfrag{y}[][][1.5]{$\frac{\delta T_r}{T_0}$}
\psfrag{x}[][][1.5]{$\frac{x}{\lambda}$}
\centering
\includegraphics[width=6.0in]{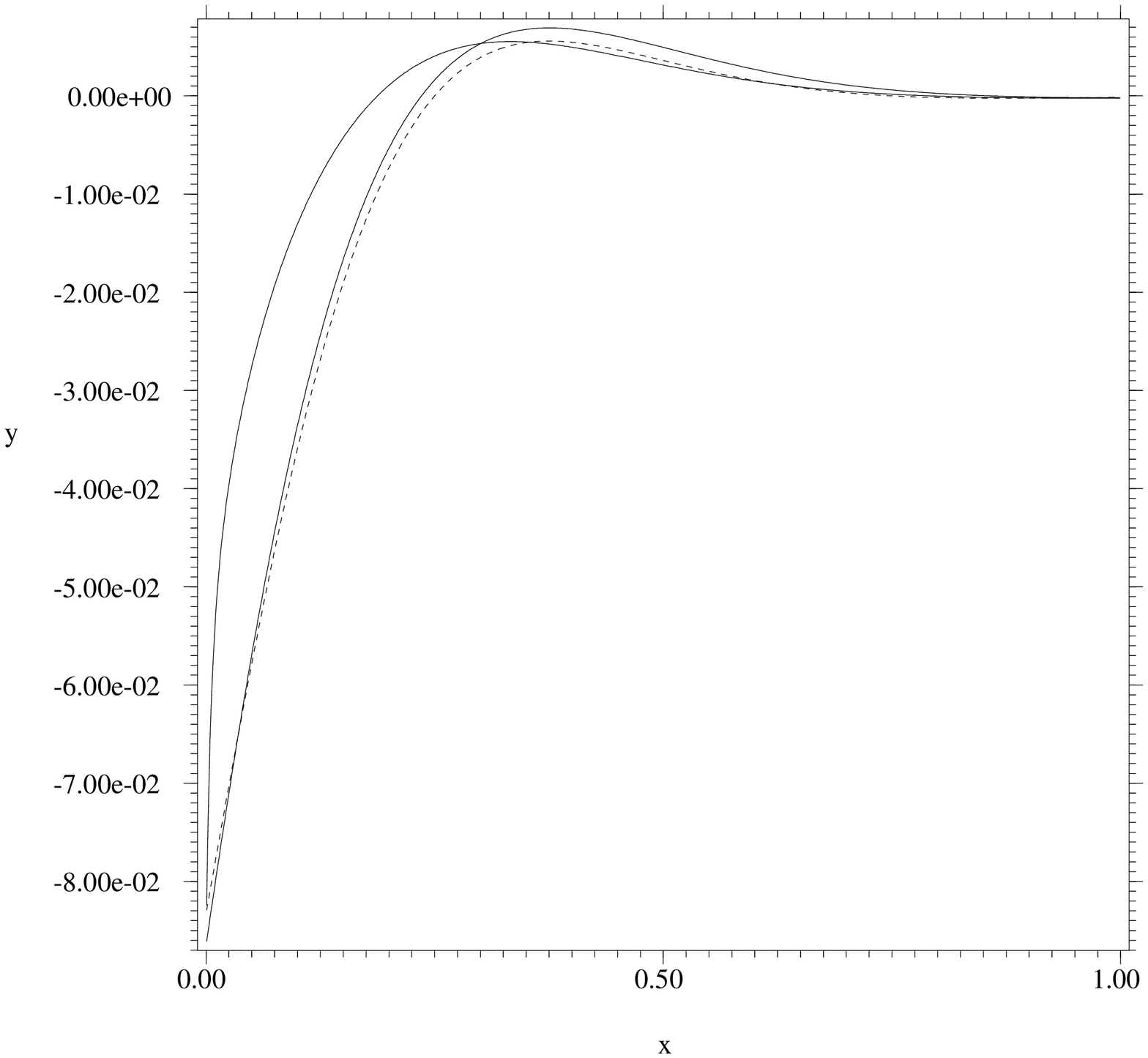}
\caption{Same as Figure~\ref{fHs1} with a higher perturbation amplitude. The differences between the numerical results with and without a flux limiter are $O(10^{-2})$.}
\label{fHs2}
\end{figure}

Some additional comments on the breakdown of flux-limited diffusion are in order. Figures~\ref{fgs1} and \ref{fgs2} are somewhat of an academic exercise, since the radiation energy in this region of parameter space exceeds the rest mass energy of the material, and our non-relativistic treatment breaks down \cite{bog96}. In addition, the superluminal propagation we have observed only occurs for waves whose amplitude is small compared to the background state, and it is not clear that this would have a significant impact on the energy budget of a realistic calculation. Finally, we have set up our computational domain to allow the excitation of any length and time scales we choose, and even with this freedom it was computationally expensive to access the regions of parameter space in which the flux limiters break down. A realistic calculation will only be able to resolve a small fraction of the length and time scales that we have explored, and only with a very large dynamic range will a calculation be able to see the excitation of superluminal waves. In any case, when computation has advanced to the point where one can easily resolve the length and time scales of free-streaming radiation, one should no longer be using flux-limited diffusion.

This work performed under the auspices of the U.S. Department of Energy by Lawrence Livermore National Laboratory under Contract DE-AC52-07NA27344.

\newpage

\clearpage

\appendix

\section{Eigenvalues and Eigenvectors}
\label{EE}

With $\partial_t \rightarrow i\omega$ and $\bnabla \rightarrow -ik$, the density, pressure and radiative flux perturbations are given by
\be\label{DRHO}
\frac{\delta \rho}{\rho_0} = \frac{\tau_c}{\tau_k} \frac{\delta v}{c},
\ee
\be\label{DP}
\frac{\delta p}{p_0} = \frac{\delta T}{T_0} + \frac{\tau_c}{\tau_k} \frac{\delta v}{c}
\ee
and
\be
\frac{\delta F}{4E_0} =  \frac{ic}{3A}\left(\tau_k^{-1} \frac{\delta T_r}{T_0} - f^\prime \tau_c^{-1}\frac{\delta v}{c} \right),
\ee
where
\be
A \equiv 1 + i f \tau_c^{-1}
\ee
and the $f$'s are flags to keep track of terms. Mihalas \& Mihalas \cite{mm83,mm84} set $f^\prime = 0$ everywhere but $f = 0$ only for the flux in the material momentum equation (otherwise $f = 1$); a consistent treatment has $f = f^\prime = 1$. For the diffusion approximation, $f = f^\prime = 0$.

Using the above expressions in the material and radiation energy equations gives 
\be
16\gamma r \left(\frac{\delta T}{T_0} - \frac{\delta T_r}{T_0}\right) = i \left([\gamma -1]\tau_k^{-1} \frac{\delta v}{c} - \tau_c^{-1}\frac{\delta T}{T_0} \right)
\ee
and
\be
\frac{\delta T}{T_0} = \left(1 + i\tau_c^{-1} + \frac{\tau_k^{-2}}{3A}\right) \frac{\delta T_r}{T_0} - i\frac{\tau_k^{-1}}{3A}\left(A - f^\prime i\tau_c^{-1}\right)\frac{\delta v}{c},
\ee
or, equivalently,
\be
C\left(\frac{\delta T}{T_0} - \frac{\delta T_r}{T_0}\right) = (C - 1)\frac{\delta T}{T_0} - i\frac{\tau_k^{-1}}{3A}\left(A - f^\prime i\tau_c^{-1}\right)\frac{\delta v}{c},
\ee
where
\be
C \equiv 1 + i\tau_c^{-1} + \frac{\tau_k^{-2}}{3A}.
\ee
Eliminating $\delta T$ and $\delta T_r$, respectively, from these equations gives
\be\label{EV1}
\left(16\gamma r [C-1] + i\tau_c^{-1} C\right)\frac{\delta T}{T_0} = i\tau_k^{-1}\left(\frac{16\gamma r}{3A}A^\prime + [\gamma - 1]C\right)\frac{\delta v}{c}
\ee
and
\be\label{EV2}
\left(16\gamma r [C-1] + i\tau_c^{-1} C\right)\frac{\delta T_r}{T_0} = i\tau_k^{-1}\left(\frac{16\gamma r + i\tau_c^{-1}}{3A}A^\prime + \gamma - 1\right)\frac{\delta v}{c},
\ee
where
\be
A^\prime \equiv 1 + (f - f^\prime) i\tau_c^{-1}.
\ee
The ratio of these expressions,
\be\label{EV3}
\frac{\delta T}{\delta T_r} = \frac{16\gamma r A^\prime + 3A(\gamma - 1)C}{(16\gamma r + i\tau_c^{-1})A^\prime + 3A(\gamma - 1)}
\ee
demonstrates that the material and radiation temperatures are nearly equal in the optically-thick limit ($\tau_c \gg 1$ and $\tau_k \gg 1$).

The material momentum equation is
\be
\tau_c\tau_k\left(\tau_k^{-2} - \gamma\tau_a^{-2} - f^\prime\frac{16\gamma r\tau_c^{-2}}{3\tilde{A}(\gamma - 1)}\right)\frac{\delta v}{c} + \frac{\delta T}{T_0} + \frac{16\gamma r}{3\tilde{A}(\gamma - 1)} \frac{\delta T_r}{T_0}
 = 0,
\ee
where
\be
\tilde{A} \equiv 1 + i\tilde{f} \tau_c^{-1}
\ee
has been defined to indicate that Mihalas \& Mihalas \cite{mm83,mm84} ignore the frequency dependent term in this equation (for them, $\tilde{f} = 0$). Replacing $\delta v$ with expression (\ref{EV1}) and $\delta T_r$ with expression (\ref{EV3}) gives the dispersion relation:
\be
c_4 \tau_k^{-4} + c_2 \tau_k^{-2} + c_0 = 0,
\ee
with
\be
c_4 = 1 - i16r\tau_c,
\ee
\bea
c_2 = 3A\left(1 + i\tau_c^{-1}\right) - \tau_a^{-2}\left(1 - i16\gamma r \tau_c\right)  \nonumber \\ + 16r\left(4A - if^\prime \tau_c^{-1} + \frac{A}{\tilde{A}}+ \frac{A}{\tilde{A}}\frac{16\gamma r + i\tau_c^{-1}}{3(\gamma - 1)}\right)
\eea
and
\be
c_0 = -3A\tau_a^{-2}\left(1 + 16\gamma r + i\tau_c^{-1}\right)\left(1 + f^\prime \frac{16ra^2}{3\tilde{A}(\gamma - 1)c^2}\right).
\ee

Under the assumptions of Mihalas \& Mihalas \cite{mm83,mm84} ($f^\prime = \tilde{f} = 0$ and $f = 1$), these become
\bea
c_2 = 3\left(1 + i\tau_c^{-1}\right)^2 - \tau_a^{-2}\left(1 - i16\gamma r \tau_c\right) \nonumber \\ + 16r\left(1 + i\tau_c^{-1}\right)\left(5 + \frac{16\gamma r + i\tau_c^{-1}}{3(\gamma - 1)}\right)
\eea
and
\be
c_0 = -3\left(1 + i\tau_c^{-1}\right)\tau_a^{-2}\left(1 + 16\gamma r + i\tau_c^{-1}\right),
\ee
with $c_4$ unchanged. These coefficients match those of expression (3.12) in Mihalas \& Mihalas \cite{mm83}. The eigenvector relationships are
\be\label{EV1MM}
\left(16\gamma r [C-1] + i\tau_c^{-1} C\right)\frac{\delta T}{T_0} = i\tau_k^{-1}\left(\frac{16\gamma r}{3} + [\gamma - 1]C\right)\frac{\delta v}{c}
\ee
and
\be\label{EV2MM}
\left(16\gamma r [C-1] + i\tau_c^{-1} C\right)\frac{\delta T_r}{T_0} = i\tau_k^{-1}\left(\frac{16\gamma r + i\tau_c^{-1}}{3} + \gamma - 1\right)\frac{\delta v}{c}.
\ee

A self-consistent treatment under the Eddington approximation has $f^\prime = \tilde{f} = f = 1$, which gives
\bea
c_2 = 3\left(1 + i\tau_c^{-1}\right)^2 - \tau_a^{-2}\left(1 - i16\gamma r \tau_c\right)  \nonumber \\ + 16r\left(5 + 3i \tau_c^{-1} + \frac{16\gamma r + i\tau_c^{-1}}{3(\gamma - 1)}\right)
\eea
and
\be
c_0 = -3\tau_a^{-2}\left(1 + 16\gamma r + i\tau_c^{-1}\right)\left(1 + i\tau_c^{-1} + \frac{16ra^2}{3(\gamma - 1)c^2}\right).
\ee
The self-consistent eigenvector relationships are
\be
\left(16\gamma r [C-1] + i\tau_c^{-1} C\right)\frac{\delta T}{T_0} = i\tau_k^{-1}\left(\frac{16\gamma r}{3\left(1 + i\tau_c^{-1}\right)} + [\gamma - 1]C\right)\frac{\delta v}{c}
\ee
and
\be
\left(16\gamma r [C-1] + i\tau_c^{-1} C\right)\frac{\delta T_r}{T_0} = i\tau_k^{-1}\left(\frac{16\gamma r + i\tau_c^{-1}}{3\left(1 + i\tau_c^{-1}\right)} + \gamma - 1\right)\frac{\delta v}{c}.
\ee

The diffusion approximation ($f^\prime = \tilde{f} = f = 0$) gives
\bea
c_2 = 3\left(1 + i\tau_c^{-1}\right) - \tau_a^{-2}\left(1 - i16\gamma r \tau_c\right)  \nonumber \\ + 16r\left(5 + \frac{16\gamma r + i\tau_c^{-1}}{3(\gamma - 1)}\right)
\eea
and
\be
c_0 = -3\tau_a^{-2}\left(1 + 16\gamma r + i\tau_c^{-1}\right).
\ee
The eigenvector relationships under the diffusion approximation are given by expressions (\ref{EV1MM}) and (\ref{EV2MM}) with $C$ replaced by
\be
C^\prime \equiv 1 + i\tau_c^{-1} + \frac{\tau_k^{-2}}{3}.
\ee

We have derived the approximate expressions (\ref{KA}) and (\ref{KD}) both analytically and by a semi-empirical approach described below. The positive and negative branches of the dispersion relation are given approximately by
\be
\tau_+^{-2} \simeq -\frac{c_0}{c_2}\left(1 + \frac{c_4c_0}{c_2^2}\right)
\ee
and
\be
\tau_-^{-2} \simeq -\frac{c_2}{c_4}\left(1 - \frac{c_4c_0}{c_2^2}\right),
\ee
where the second term in parentheses is only required to derive the expressions in regions a and c. Asymptotic expansions are performed on the coefficients of the dispersion relation and the leading terms are inserted in the above expressions. As it turns out, these asymptotic expansions are not always easy to perform, and we have been unable to derive the approximate expression in region g analytically.\footnote{An important consideration with the analytical approach is to only perform asymptotic expansions of complex expressions that appear in the numerator; complex expressions in the denominator should be converted to real expressions via their complex conjugate.} A more straightforward semi-empirical approach is to calculate linear fits of the full solutions on a log-log plot to determine the scaling of the solutions with the various parameters. We have done this to obtain the expression for region g as well as to check our analytical approach for the other regions.

\section{Details on the Vincenti \& Baldwin Analysis}
\label{VB}

Equation (51) of Vincenti \& Baldwin \cite{vb62} is
\bea
i \frac{\nbo}{8\gamma\nbu}\left(H[\xi] + H^{\prime\prime}[\xi]\right) = \left(B - H[0] - \gamma^{-1} H^{\prime\prime}[0]\right)E_2(\nbu\xi) \nonumber \\ \; - \int_0^\xi E_2\left(\nbu[\xi - \txi]\right) \left(H^{\prime}[\txi] + \gamma^{-1} H^{\prime\prime\prime}[\txi]\right)d\txi \nonumber \\ \; + \int_\xi^\infty E_2\left(\nbu[\txi - \xi]\right) \left(H^{\prime}[\txi] + \gamma^{-1} H^{\prime\prime\prime}[\txi]\right)d\txi,
\eea
where $H$ is a dimensionless perturbation amplitude, $B$ is a boundary condition, a prime denotes a derivative with respect to $\xi$ (a dimensionless spatial coordinate) and
\be
E_2(z) \equiv \int_0^1 e^{z/s} ds.
\ee
At this point Vincenti \& Baldwin \cite{vb62} replace $H$ with $\sj C_j e^{c_j \xi}$ (a sum of exponentials) and approximate $E_2$ with a single exponential. This approximation is unnecessary, however. Making only the former substitution gives
\bea
i \frac{\nbo}{8\gamma\nbu} \sj \left(1+ c_j^2\right) C_j e^{c_j \xi} = \left(B - \sj C_j \left[1 + \gamma^{-1} c_j^2 \right] \right)E_2(\nbu\xi) \nonumber \\ \; -  \int_0^\xi  d\txi \int_0^1 d\mu \; e^{-\nbu \xi/\mu + \nbu \txi/\mu} \sj c_j \left(1 + \gamma^{-1} c_j^2\right) C_j e^{c_j \txi}\nonumber \\ \; + \int_\xi^\infty d\txi \int_0^1 d\mu \; e^{\nbu \xi/\mu - \nbu \txi/\mu} \sj c_j \left(1 + \gamma^{-1} c_j^2\right) C_j e^{c_j \txi}.
\eea
Changing the order of integration, taking the $\xi$ factor out of the integral over $\txi$ as well as the constants out of both integrals, and reversing the limits of integration on the second integral over $\txi$ gives
\bea
i \frac{\nbo}{8\gamma\nbu} \sj \left(1+ c_j^2\right) C_j e^{c_j \xi} = \left(B - \sj C_j \left[1 + \gamma^{-1} c_j^2 \right] \right)E_2(\nbu\xi) \nonumber \\ \; - \sj c_j \left(1 + \gamma^{-1} c_j^2\right) C_j \left( \int_0^1 d\mu \; e^{-\nbu \xi/\mu} \int_0^\xi  d\txi \; e^{(c_j+\nbu/\mu)\txi}\nonumber \right. \\ \left. \; + \int_0^1 d\mu \; e^{\nbu \xi/\mu} \int_\infty^\xi d\txi \; e^{(c_j-\nbu/\mu)\txi}\right).
\eea
Performing the integrals over $\txi$ gives
\bea\label{VB51}
i \frac{\nbo}{8\gamma\nbu} \sj \left(1+ c_j^2\right) C_j e^{c_j \xi} = \left(B - \sj C_j \left[1 + \gamma^{-1} c_j^2 \right] \right)E_2(\nbu\xi) \nonumber \\ \; - \sj c_j \left(1 + \gamma^{-1} c_j^2\right) C_j e^{c_j\xi} \int_0^1 d\mu \; \left[\frac{\mu}{c_j\mu+\nbu} + \frac{\mu}{c_j\mu-\nbu} \right]
\nonumber \\ \; + \sj c_j \left(1 + \gamma^{-1} c_j^2\right) C_j \int_0^1 d\mu \; \frac{\mu}{c_j\mu+\nbu} e^{-\nbu \xi/\mu},
\eea
where $Re(c_j) < 0$ has been assumed (this is necessary for the perturbations to remain finite as $\xi \rightarrow \infty$). The integral over $\mu$ in the second line above is
\be
\int_0^1 d\mu \; \frac{2c_j\mu^2 }{c_j^2\mu^2-\nbu^2} = \frac{2}{c_j}\left[1 - \frac{\nbu}{c_j}\tanh^{-1}\left(\frac{c_j}{\nbu}\right)\right].
\ee

The terms proportional to $e^{c_j\xi}$ in expression (\ref{VB51}) yield the dispersion relation\footnote{This is equivalent to expression (\ref{VBDR}) in the text after making the substitutions (\ref{VBC1})-(\ref{VBC3}).}
\be\label{VBDR2}
1+ c_j^2 - i \frac{16 \nbu} {\nbo} \left(\gamma + c_j^2\right) \left[1 - \frac{\nbu}{c_j}\tanh^{-1}\left(\frac{c_j}{\nbu}\right)\right] = 0,
\ee
while the remainder of the terms imply
\be\label{VBBC}
\int_0^1 d\mu \;  e^{-\nbu \xi/\mu} \left[B - \sj C_j \left(1 + \gamma^{-1} c_j^2 \right)\frac{\nbu}{c_j\mu+\nbu} \right] = 0.
\ee
This expresses the boundary condition at the wall. It can be rewritten in terms of exponential integrals and integrated, but it does not appear that this expression can be satisfied for all $\xi$. This may be due to the fact that the discrete spectrum is not sufficient to match the particular boundary condition chosen by Vincenti \& Baldwin \cite{vb62}. Equation (\ref{VBDR2}) also gives rise to a continuous spectrum of modes when the wavenumber is purely imaginary and the optical depth is unity ($c_j/\nbu = \pm 1$).\footnote{This corresponds to a singularity in the dispersion relation when the argument of $\tanh^{-1}$ is $\pm 1$. See Bogdan et al. \cite{bog96} for further discussion.} Including the continuous spectrum in the decomposition of $H(\xi)$ would likely alleviate the discrepancy in equation (\ref{VBBC}).

\end{document}